\documentclass{aa}  
\usepackage{graphicx}
\usepackage{txfonts}
\def\lsim{\mathrel{\rlap{\lower 3pt \hbox{$\sim$}} \raise 2.0pt \hbox{$<$}}}
\def\gsim{\mathrel{\rlap{\lower 3pt \hbox{$\sim$}} \raise 2.0pt \hbox{$>$}}}
\newcommand{\msun}{\ensuremath{M_\odot}}
\newcommand{\msf}{\ensuremath{M_{\rm SF}}}
\newcommand{\td}{\ensuremath{t_{\rm D}}}
\newcommand{\tu}{\ensuremath{t_{\rm U}}}
\newcommand{\kia}{\ensuremath{k_{\rm Ia}}}
\newcommand{\kcc}{\ensuremath{k_{\rm CC}}}
\newcommand{\fia}{\ensuremath{f_{\rm Ia}}}
\newcommand{\taun}{\ensuremath{\tau_{\rm n}}}
\newcommand{\taunx}{\ensuremath{\tau_{\rm n,x}}}
\newcommand{\taugr}{\ensuremath{\tau_{\rm GW}}}

\begin{document} 

   \title{Correlation of the rate of Type Ia supernovae with the parent galaxy properties: Light and shadows}
   \titlerunning{SNIa rate and the properties of the parent galaxy}

   \author{L. Greggio\inst{1}\thanks{email:laura.greggio@inaf.it},
         E. Cappellaro
          \inst{1}
          }

   \institute{INAF - Osservatorio Astronomico di Padova,
             Vicolo dell'Osservatorio 5, Padova, 35122 Italy\\
             }

      \date{}
 
\abstract
{The identification of the progenitors of Type Ia supernovae (SNIa) is extremely important in several astrophysical contexts, ranging from   stellar evolution in close binary systems to evaluating cosmological parameters. Determining the distribution of the delay times (DTD) of SNIa progenitors can shed light on their nature. The DTD can be constrained by analysing the correlation between the SNIa rate and those properties of the parent galaxy which trace the average age of their stellar populations.} 
{We investigate the diagnostic capabilities of this correlation by examining its systematics with the various parameters at play: simple stellar population models, the adopted description for the star formation history (SFH) in galaxies, and the way in which the masses of the galaxies are evaluated.}
{We computed models for the diagnostic correlations for a variety of  input ingredients and for a few astrophysically motivated DTD laws appropriate for a wide range of possibilities for the SNIa progenitors. The models are compared to the results of three independent observational surveys.}
{The scaling of the SNIa rate with the properties of the parent galaxy is sensitive to all input ingredients mentioned above. This is a severe limitation on the possibility to discriminate alternative DTDs. In addition, current surveys show some discrepancies for the reddest and bluest galaxies, likely because of  limited statistics and the inhomogeneity of the observations. For galaxies with intermediate colours the rates are in agreement, leading to a robust determination of the productivity of SNIa from stellar populations of $\simeq$ 0.8 events per 1000 \msun.}  
 {Large stastistics of SNIa events along with accurate measurements of the SFH in the galaxies are required to derive firm constraints on the DTD. The LSST will achieve these results by providing  a homogeneous, unbiased, and vast database on both SNIa and galaxies.}
   {}


   \maketitle
%

\section{Introduction}

The identification of the progenitors of supernovae of Type Ia (SNIa) is of great importance in several astrophysical contexts, such as constraining the evolutionary paths of close binary systems \citep[e.g.][]{Claeys_2014}, measuring the cosmological parameters \citep{riess:1998qm,Perl_1999}, studying the chemical evolution of galaxies \citep{matteucci:1986kq,TGB_1999,Kobayashi_2015} and of the intergalactic medium \citep{Maoz_2004,blanc:2008rp}, the evolution of galactic winds \citep{Ciotti_1991}, and modelling the gravitational waves emission from binary stars \citep{Farmer_2003}. Yet, in spite of great theoretical and observational efforts in recent decades, the question is still far from settled \citep{MM_2012,Maeda_2016}. While there is general consensus that the events originate from the thermonuclear explosion of a white dwarf (WD) member of a close binary system, there are many possible evolutionary paths leading to a successful explosion, each involving different ranges of initial masses and separations, and different delay times, that is the time between the formation of the primordial binary and the SNIa explosion. In general, the evolutionary paths can be accommodated in two classes, depending on the nature of  the companion of the exploding WD. If the companion is an evolving star the evolutionary path is called single degenerate  
\citep[SD;][]{whelan:1973nr}; if the companion is another WD, the path is called double degenerate 
\citep[DD;][]{Webbink_1984,IT_1984}.  In both scenarios the exploding star is a carbon-oxygen (CO) WD which 
ignites nuclear fuel under degenerate conditions, following accretion from a close companion. Explosion may occur either because the CO WD manages to reach the Chandrasekhar limit, or because a sufficiently massive helium layer has accumulated on top of the WD and detonates.
The diversity of SNIa light curves indicates that both kinds of explosions occur in nature see \citep[see][for a comprehensive review]{Hillebrandt_2013}. In addition, the correlation of the light curve characteristics with the host galaxy properties suggest that different kinds of events may pertain to different ages or metallicities of the parent stellar population \citep[e.g.][]{Childress_2013}.

Independent of the explosion mechanism, the delay time essentially reflects the evolutionary channel. While for SD progenitors accretion starts when the secondary component of the binary evolves off the stellar main sequence, 
 in the DD scenario at this point of the evolution a common envelope (CE) forms around the two stars and is eventually expelled from the system, hindering the mass growth of the WD (however see \cite{Ruiter_2013} for a more detailed view). During the CE phase, the system looses orbital energy and shrinks, so that the final configuration consists of two WDs with a separation smaller than that at the beginning of the CE phase. Hereafter, the DD system looses angular momentum because of gravitational waves radiation, and eventually the two WDs merge.  Traditionally these two channels are considered in mutual competition, but there is evidence which supports that both paths are at work \citep[][and references therein]{patat:2007sh,GRD08,Li_2011Nature,gonzales_2012,cao_2015,Hosseinzadeh_2017,Dimitriadis_2019}. Information on the relative contribution of the various channels requires collecting big samples of data to be able to relate the properties of individual explosions to the specific evolutionary path, as well as to build up a significant statistics. A less ambitious goal consists in determining the distribution of the delay times of SNIa explosions (DTD), which is proportional to the time evolution of the SNIa rate following an instantaneous burst of star formation (SF). Indeed, the DTD  provides clues on the evolutionary channels of SNIa progenitors, insofar as they correspond to different proportions of events at different delay times. In addition, the DTD is interesting in itself because it regulates the energetic and chemical input in an evolving stellar system, and is thus a fundamental ingredient to model the evolution of a variety of astrophysical objects, for example star clusters, galaxies, clusters of galaxies, and interstellar and intergalactic medium. 
 
Population synthesis codes provide theoretical renditions of the DTD which include a variety of different channels following from the predictions of stellar evolution in close binary systems \citep[e.g.][]{Ruiter_2009,mennekens_2010,wang_2012,Claeys_2014}. These models offer a comprehensive scenario of the evolutionary outcome, but their results are sensitive to many assumptions, some of which regard poorly constrained astrophysical processes. Particularly critical is the description of  the CE phase and especially its efficiency in shrinking the system. In this respect we notice that to produce merging within a Hubble time, the separation of the DD system should be reduced down to a few solar radii or $\sim$ few percent of its value at the first Roche Lobe overflow \citep[cf.][]{greggio_2010}.

Alternatively, the DTD can be derived empirically from the analysis of the relation of the SNIa rate with the properties of the parent stellar population 
\citep[e.g.][]{mannucci:2006zi,scannapieco:2005rr}. This can be understood considering the following equation which links the supernova rate at a time $t$, $\dot{n}_{\rm Ia}(t)$,  to the DTD, $f_{Ia}(\td)$  :

\begin{equation}
\dot{n}_{\rm Ia}(t) = \kia \int_0^t {\psi(t-\td)\,\fia (\td)\,\rm{d} \td}
\label{eq_rate}
,\end{equation}

\noindent
where $\td$ is the delay time, $\psi$ is the star formation rate (SFR), and $\kia$ is the number of SNIa per unit mass of one stellar generation, usually assumed constant in time \citep[cf.][]{greggio_2005,greggio_2010}.
Following Eq. (\ref{eq_rate}), the SNIa rate in a galaxy results from the convolution of the DTD and the past star formation history (SFH) over the whole galaxy lifetime, since the delay times of SNIa progenitors span a wide range, possibly up to the Hubble time and beyond. Provided that the DTD is not flat, Eq. (1) implies a correlation of the SNIa rate with the shape of the SFH, modulated by the DTD. It is then possible to derive information on the DTD by analysing the relation between the SNIa rate and those properties of the parent galaxy which trace the SFH. 
Since the occurrence of a supernova is a rare event, the correlations are usually constructed by considering large samples of galaxies and averaging the rates and galaxy properties in wide bins.
To do this we need to scale the SNIa rate according to a parameter which describes the galaxy size, for example the total luminosity or total mass. Depending on the chosen parameter, the slope of the correlation changes, so that, for example the trend of the SNIa rate with the colour of the parent galaxy is flatter when normalizing the rate to the total B-band luminosity rather than to the K-band luminosity \citep{greggio+2009} or to the galaxy mass \citep{Mannucci_2005}.  From Eq. (\ref{eq_rate}) it is clear that the most direct normalization is obtained scaling the SNIa rate to the integral of the SFH over the galaxy lifetime. 

Many different observed correlations have been considered in the literature to derive information on the DTD: (i) the rate per unit mass (and/or luminosity) trend with the colour (and/or morphological type) of the parent galaxy \citep{greggio_2005,mannucci:2006zi}; (ii) the rate per unit mass dependence on the specific SFR (sSFR) \citep{pritchet_2008}; and (iii) the rate per unit mass versus the mass of the parent galaxy \citep{graur:2013fd}. Correlations (i) and (ii) have been widely interpreted as showing that  the rate depends on the age distribution of the stars in the parent galaxy. Correlation (iii) may indirectly reflect the same property through the observed galaxy downsizing phenomenon \citep[e.g.][]{gallazzi_2005}: the SNIa rate is expected to be higher in the less massive galaxies because of the younger average age of their inhabiting stellar populations compared to the stars in the more massive galaxies \citep{graur:2013fd}. All observed correlations are compatible with the notion that the rate per unit mass is higher in younger systems, while the rendition of this  can be described through different tracers of the average stellar age. This supports the notion of a DTD which is more populated at the short delay times \citep{greggio_2005}.

In order to analyse the physics behind the above empirical correlations we need to specify the galaxy SFH. So far, most works in this field have adopted some analytic description of the SFH, for example an exponentially declining, a delayed exponential, or a power law, which all include an encoded free parameter. 
By varying this parameter and the age of the model, both the predicted galaxy properties (mass, colours, current SFR, etcetera) and SNIa rate change and their correlation is modulated by the DTD \citep[e.g.][]{greggio_2005,mannucci:2006zi,pritchet_2008}.
By comparing the predicted correlation to the observations, the shape of the DTD remains constrained. 

\cite{maoz+2011} adopted a different approach, in which individual galaxies are examined with the VESPA code to derive their stellar age distribution in a few wide bins by fitting their spectral energy distribution (SED), while the DTD is described as a generic power law. Depending on an assumed value of the power law exponent, a probability of occurrence of a SNIa event in each galaxy is computed and the analysis of the statistical properties of the sample leads to the determination of a best fit DTD slope. This method presents the advantage of avoiding averaging the galaxy  properties in wide bins and traces the capability of each galaxy to produce SNIa events. On the other hand, it requires an accurate knowledge of the SFH in each galaxy, which is hard to achieve. Also, there is no guarantee that the global SFH which results from the galaxy mixture reproduces the observed cosmic SFH. In contrast, the general correlations between observed quantities reflect global properties of the stellar populations, averaging out the individual evolutionary paths followed by each galaxy.

In all these cases the SFH is constrained by the galaxy SED in one way or another. More recently, in the context of galaxy evolution studies  a novel method has been developed. In this approach, the SFH law is constrained by the observed global properties  of  large samples of galaxies,  for example the correlation between the SFR and galaxy mass, and its evolution with redshift; the existence of a dichotomy in the colour of galaxies, which populate the blue cloud and the red sequence; and the evolution of the cosmic SFR density \citep{Renzini_2009,Peng_2010,Oemler_2013,renzini_2016,chiosi_2017,Eales_2018}.  These properties can be arranged to construct a consistent picture of the general features of the evolution of the SFR in galaxies over cosmic times. We consider two formulations of the SFH obtained under this view \citep{Peng_2010,Gladders_2013}, both of which  describe well the evolution with redshift of the galaxy luminosity functions \citep{Peng_2010,abramo_2016}. In contrast to the SED fitting of individual galaxies, this approach aims at accounting for the general behaviour of the SFH of galaxies in the evolving universe,
and are well suited to model the trend of the SNIa rate with galaxy properties, which is necessarily measured in big galaxy samples. In the following we address this kind of description of the SFH as `` cosmological'', to distinguish it  from the other, more `` standard'' approach, which considers each galaxy independently.
In this paper we show that the choice of the SFH description has an impact on the predicted correlations and hence it adds a significant contribution to the systematic uncertainties that has been neglected or underestimated so far.

In \cite{Botticella_2017} we discussed the constraints on the DTD which could be obtained from the correlations of the SNIa rate with the parent galaxy properties found in the SUDARE (Supernova Diversity and Rate Evolution) survey \citep{cappellaro+2015}, adopting a standard description of the SFH.   
In this paper we expand on those results, including the cosmological SFH laws, as well as a comparison with more observational data, i.e. the rates measured in local galaxies from the LOSS (Lick Observatory Supernova Search) search \citep{Li_2011,Graur_2017} and from the \cite{cappellaro:1999dg}  search (hereafter CET99). 
For the three surveys we use the original SN discovery list \footnote{We notice that for the LOSS survey the detailed SN subtype classification is available, but we do not consider this classification. For our analysis we separate the SNe in two main groups: type Ia and core collapse.}  and control times to compute supernova rates for different galaxy parameters and binning with respect to the published values. This allows us to optimize the description of the trend of the rate with the properties of the parent galaxy, as well as the comparison of the results from the three independent datasets. For the LOSS survey we use the control times published by \cite{Graur_2017}, while the SUDARE and CET99 control times were already available to us. We checked that we recover the results published in \cite{Li_2011} and \cite{Mannucci_2005}  when using the same parameters and binning. Using the original data for the surveys we were also able to test various approaches to estimate galaxy masses (cf. Sect. 5).
We remark that the SUDARE survey targets galaxies in the  Chandra Deep Field South (CDFS) and in the Cosmic Evolution Survey (COSMOS) fields and considers only objects with redshift  $z \leq 1$. The vast majority of galaxies in the SUDARE sample are at intermediate redshift ($z \gtrsim 0.3$) and their redshift distribution does not show any prominent peak. 
We focus on the correlations of the SNIa rate with the parent galaxy colours, and along with different descriptions of the SFH, we test different sets of simple stellar population  models to compute galaxy colours, which impact on this kind of analysis. 

 The paper is organized as follows. In Section 2 we describe the ingredients of the computations of the theoretical correlations between the SNIa rate and the parent galaxy colours (DTD and SFH). The model colours are compared to the data of the SUDARE galaxy sample in Section 3 to test the adequacy of our approach. In Section 4 we present the expected correlation of the SNIa rate with the $U-J$ colour to illustrate the interplay between the DTD and the SFH.  In Section 5 we present the observational scaling of the SNIa rate with the colours of the parent galaxy and discuss the effect of the assumptions on the galaxy mass-to-light ratio. In Section 6 we illustrate the diagnostic capabilities of these correlations by comparing a set of models to literature data; in Section 7 we summarize the results and draw some conclusions concerning future surveys.

\section{Model ingredients}

\subsection{Distribution of delay times}

For the DTD we adopt a selection of models from \cite{greggio_2005}. These analytic formulations, described below, are based on general arguments which take into account the clock of the explosions and  the range of initial masses of the stars in systems which can provide successful explosions. For comparison we also consider an empirical DTD akin to the results in \cite{Totani_2008} and \cite{Maoz_2012}, which we describe as a power law with an index of $-1$ from a minimum delay time of 40 Myr  (hereafter indicated with P.L. model).
We notice that by construction this DTD is not motivated by a specific astrophysical scenario and the physical interpretation of the results is left to subsequent analysis.

\begin{figure}
\centering
\resizebox{\hsize}{!}{
\includegraphics[angle=0,clip=true]{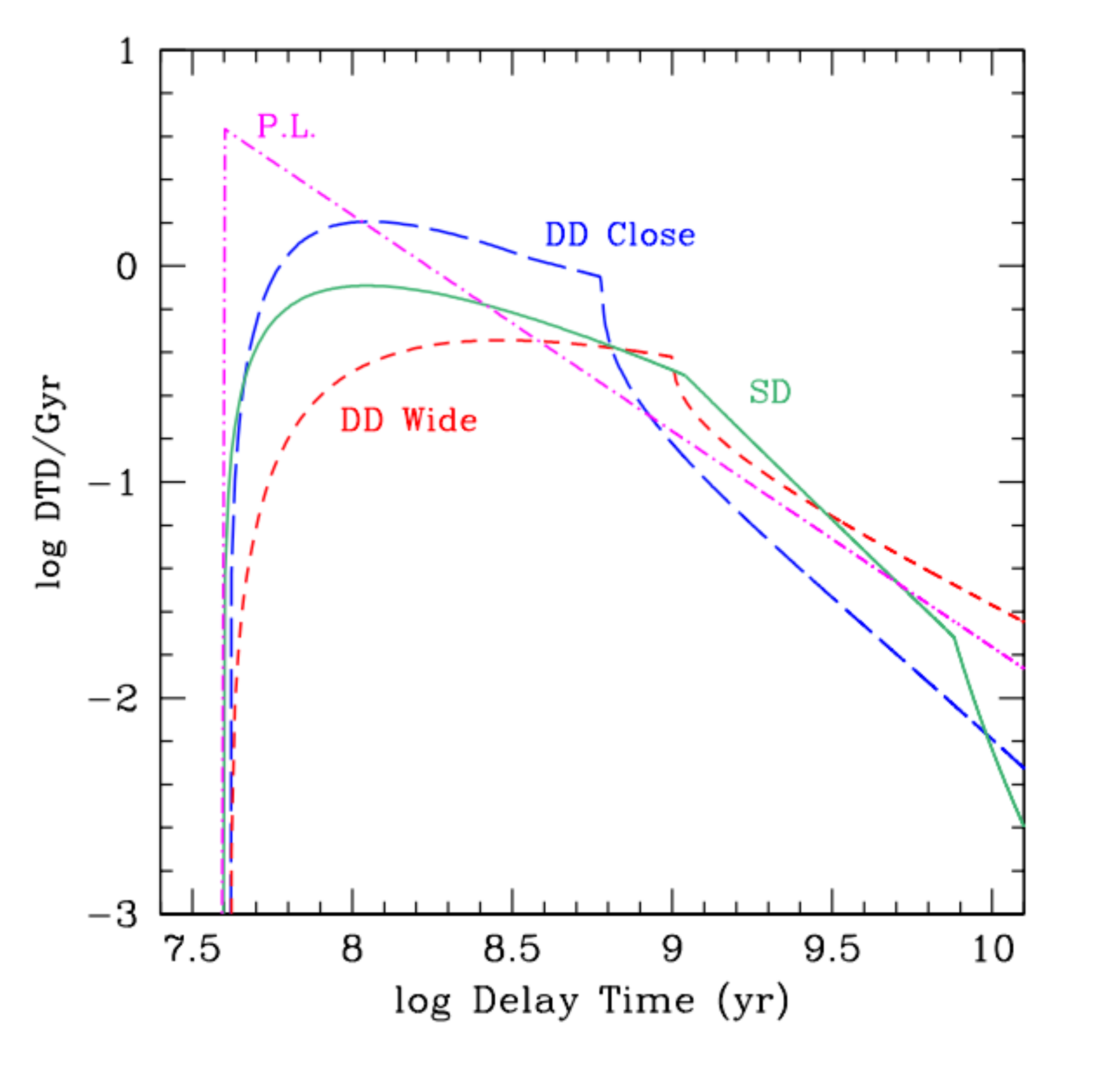}}
   \caption{Distribution of delay times for SD (solid line) and DD models (DDC, long dashed; DDW; short dashed) used in this paper. Because of the choice of parameters (see text) the DDC DTD accommodates a large fraction of prompt events, while  the DDW DTD is relatively flat. The dash-dotted line is a a power law $\propto \td^{-1}$. All DTDs considered are equal to zero at delay times shorter than 40 Myr (the evolutionary lifetime of a $\simeq$ 8 \msun\ star), and are normalized to 1 in the range 40 Myr $\leq \td \leq$ 13 Gyr.}
\label{fig_dtds}
\end{figure}

For the analytic models we consider three options:  the single degenerate (SD) model, and two versions of the double degenerate progenitor model, labelled DD wide (DDW) and DD close (DDC) (cf. Fig. \ref{fig_dtds}). As mentioned above, it is likely that SNIas arise from both single and double degenerate progenitors; the DTD of mixed models is expected to be intermediate between that of single models depending on the fractional contribution of the two components. Since we want to explore the possibility of discriminating specific channels, in this paper we do not consider mixed models.

In the SD model the time delay is virtually equal to the evolutionary lifetime of the secondary in the core hydrogen burning phase (\taun) so that early explosions are provided by systems with more massive secondaries, while late epoch events occur in systems with low mass secondaries. There are two discontinuities in the DTD of the SD model (cf. Fig. \ref{fig_dtds}) due to requirements on the mass of the progenitor: the first, at \td = 1 Gyr, comes from the lower limit of 2 \msun\, for the mass of the primary, since stars with lower mass in a close binary are more likely to produce a He, rather than a CO, WD. The second discontinuity, at \td $\simeq$ 8 Gyr, comes from the requirement that the WD mass and the envelope mass of the evolving secondary sum up to the Chandrasekhar limit. At this age the mass of the secondary is $\sim 1.1 \msun$ and the envelope mass is $\lesssim 0.8 \msun$. In order to reach the Chandrasekhar limit the accreting WD must be more massive than $\sim$ 0.6 \msun, the descendant of a $\sim 2 \msun$ primary. At later delay times, the lower mass secondary has a smaller envelope, which demands a primary more massive  than 2 \msun\  to meet the Chandrasekhar limit. 
The limitation on the primary masses in the progenitor systems, whose range progressively shrinks as the secondary mass decreases,  causes the rapid drop of the DTD at late delay times which characterizes the SD model. We notice that the delay time at which this limitation sets in depends on the fraction of envelope mass  of the secondary which is actually accreted and burned on top of the CO WD.  The SD model DTD in Fig. \ref{fig_dtds} assumes that all of the envelope of the secondary is accreted and burned on top of the companion; if only a fraction of the envelope of the secondary is used to grow the WD,  the minimum primary mass in successful systems becomes larger, and the drop sets it at earlier delay times \citep[see][]{greggio_2010}.
 
In the DD model the delay time is the sum of the evolutionary lifetime of the secondary (\taun) plus  the time it takes to the gravitational wave radiation to shrink the system up to merging of the DD components (\taugr). In \citet{greggio_2005} only the double CO WD channel is considered to lead to a successful SNIa explosion, which implies an upper limit of $\sim 1$ Gyr to \taun. The value of \taugr\ is very sensitive to the initial separation of the double degenerate system ($A$), which results after one or more CE phases. 
We point out that to merge within a Hubble time, the separation of the DD system at birth  must be smaller than a few solar radii, while the separation of the primordial binary should be at least of  a few tens of solar radii to avoid premature merging and allow the formation of a CO WD from the primary. Therefore, a high degree of shrinkage is necessary to construct a successful SNIa progenitor in the close binary evolution.
If the process is such that the more massive the binary the more it shrinks, there is little room for massive systems to explode on long delay times: both their delays \taun\ and \taugr\ are short. In this case (DDC) at long delay times the DTD  is populated by the less massive systems which manage to keep a large enough separation when they emerge from the CE phase. Conversely, if there is no dependence of the shrinkage from the binary mass, massive systems can also emerge from the CE with large values of $A$, and the DTD in this case (DDW) is flatter compared to the DDC option. The DTDs for the DD models are characterized by a plateau at short delay times, followed by a (close to a) power law decline. In these analytic  formulations, two parameters control the DTD of the DD models: the lower limit to the secondary mass in SNIa progenitors, which corresponds to an upper limit to the lifetime of the secondary (\taunx), and the distribution of the separations of the DD systems at birth, arbitrarily described as a power law. At a delay time equal to \taunx\ the DTD presents a cusp due to the setting in of a sharp upper limit to the evolutionary lifetime of the secondary, so that at delay times $\td > \taunx$ there is a lower limit to  $\taugr (= \td-\taunx),$ which increases with \td\ increasing. Thus, the minimum mass of the secondary in SNIa progenitors controls the width of the DTD plateau. The slope of the decline at delay times longer than \taunx\ is instead controlled by the distribution of the separations: the steeper this distribution, the higher the fraction of systems with short \taugr, and therefore the steeper the decline of the DTD.

In spite of the approximations introduced to derive these analytic DTDs, they compare well with the results of population synthesis codes, when taking into account the appropriate mass ranges and kind of progenitors \citep{greggio_2005, greggio_2010}. By construction, these analytic formulations provide a means to explore how the predicted rates and correlations change  when varying the DTD
under astrophysically motivated arguments concerning  the masses, separations, and their distributions, of the progenitor systems. 

Fig. \ref{fig_dtds} shows the four DTD distributions at the basis of the computations presented in this paper. Besides the empirical law
and SD model discussed above, we consider a DDC model in which the minimum mass of the secondary in the progenitor systems is of 2.5 \msun, implying \taunx = 0.6 Gyr, and a distribution of the separations of the DD systems $n(A)\propto A^{-0.9}$, i.e. close to that of unevolved binary systems \citep{Kouwenhoven_2007} . The selected DDW model, instead, adopts a minimum mass of the secondary of 2 \msun, and a flat distribution of the separations $n(A) \propto A^{0}$. These values encompass a wide range of possibilities for the progenitor systems.

\begin{figure}
\centering
\resizebox{\hsize}{!}{
\includegraphics[angle=0,clip=true]{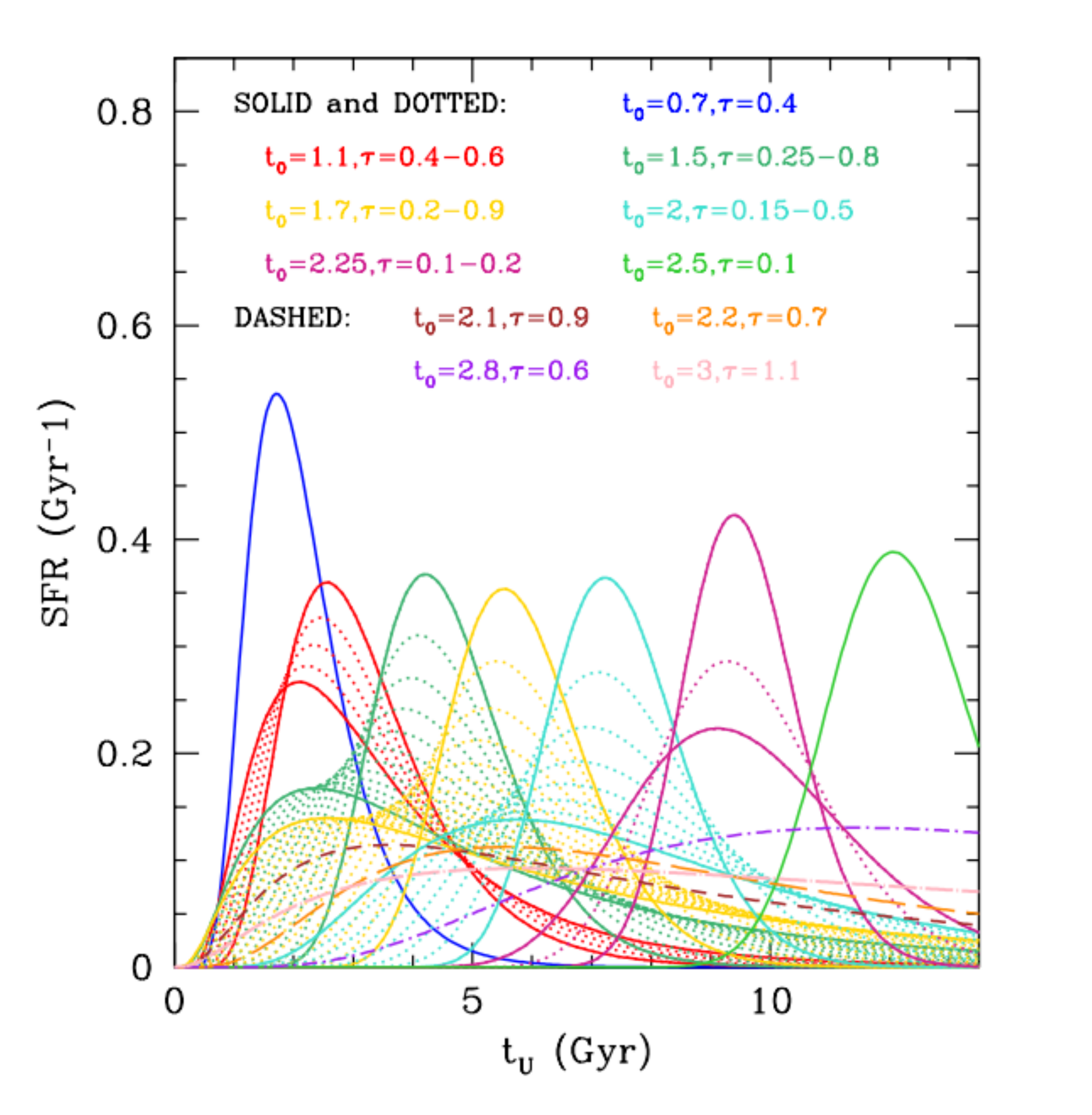}}
 \caption{Selection of $\psi(t)$ functions used to compute the models under the cosmological SFH proposed in Gladders et al. (2013).
The colour encodes the values of the parameters $t_0$ and $\tau$ as labelled. Solid and dotted lines of the same colour share the value of $t_0$, but differ in $\tau$, which is varied between a minimum and a maximum value in steps of 0.05 Gyr.
 The four curves plotted as dashed lines describe galaxies which fall outside the most populated region in Fig. 9 of Gladders et al. All time variables are in gigayear. For each $\psi(t)$ function, the maxima occur when the age of the Universe is $t_{\rm U} = {\rm exp}(t_0-\tau^2)$.}
\label{fig_abramo}
\end{figure}

\begin{figure}
\centering
\resizebox{\hsize}{!}{
\includegraphics[angle=0,clip=true]{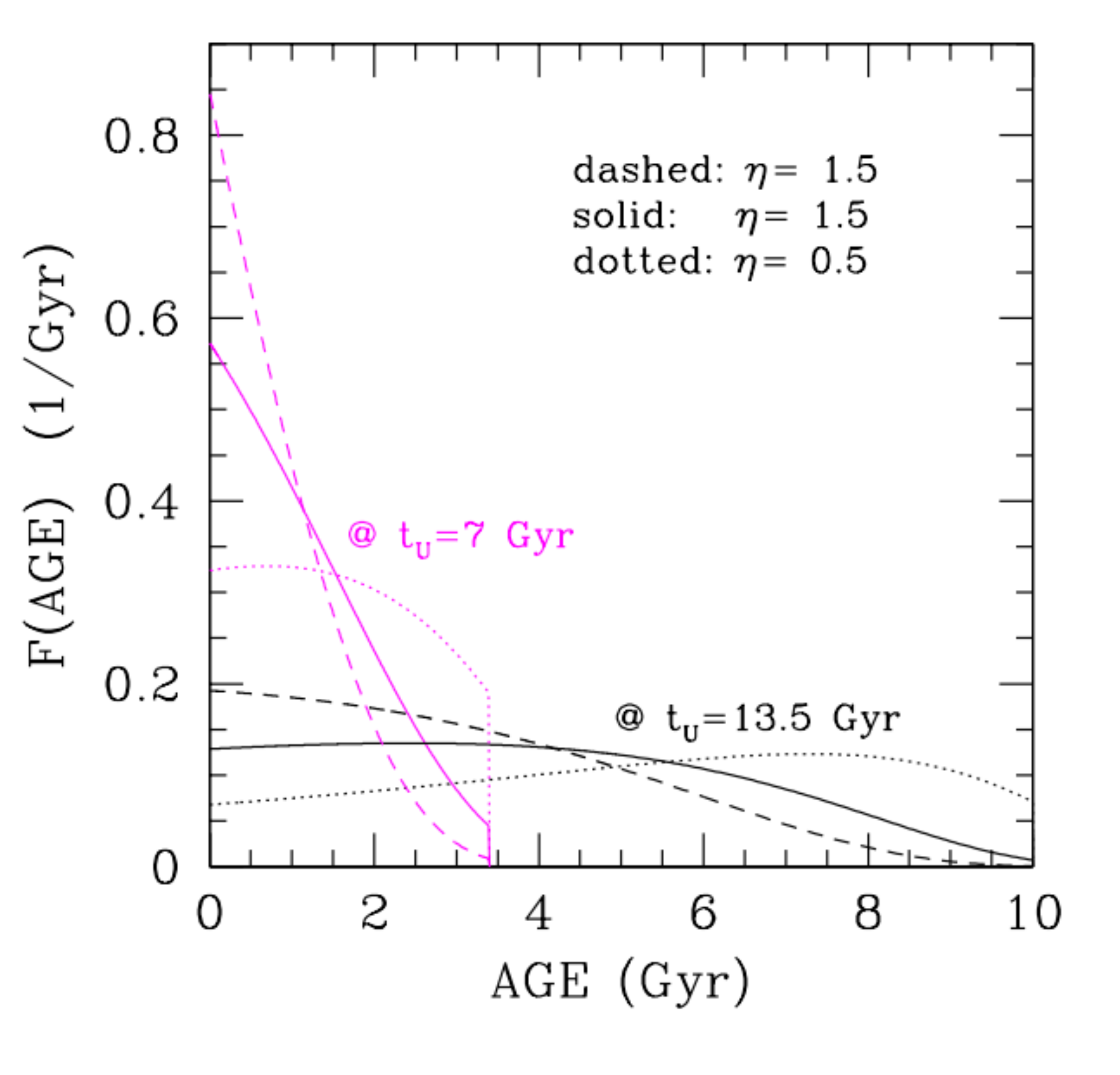}}
   \caption{Age distribution for the inflating SFH model at 2 cosmic epochs: 7 Gyr (magenta) and 13.5 Gyr (black). The three line types  refer to the three values for the parameter $\eta$, as labelled.}
\label{fig_peng}
\end{figure}

\begin{figure*}
\centering
\resizebox{\hsize}{!}{
\includegraphics[angle=0,clip=true]{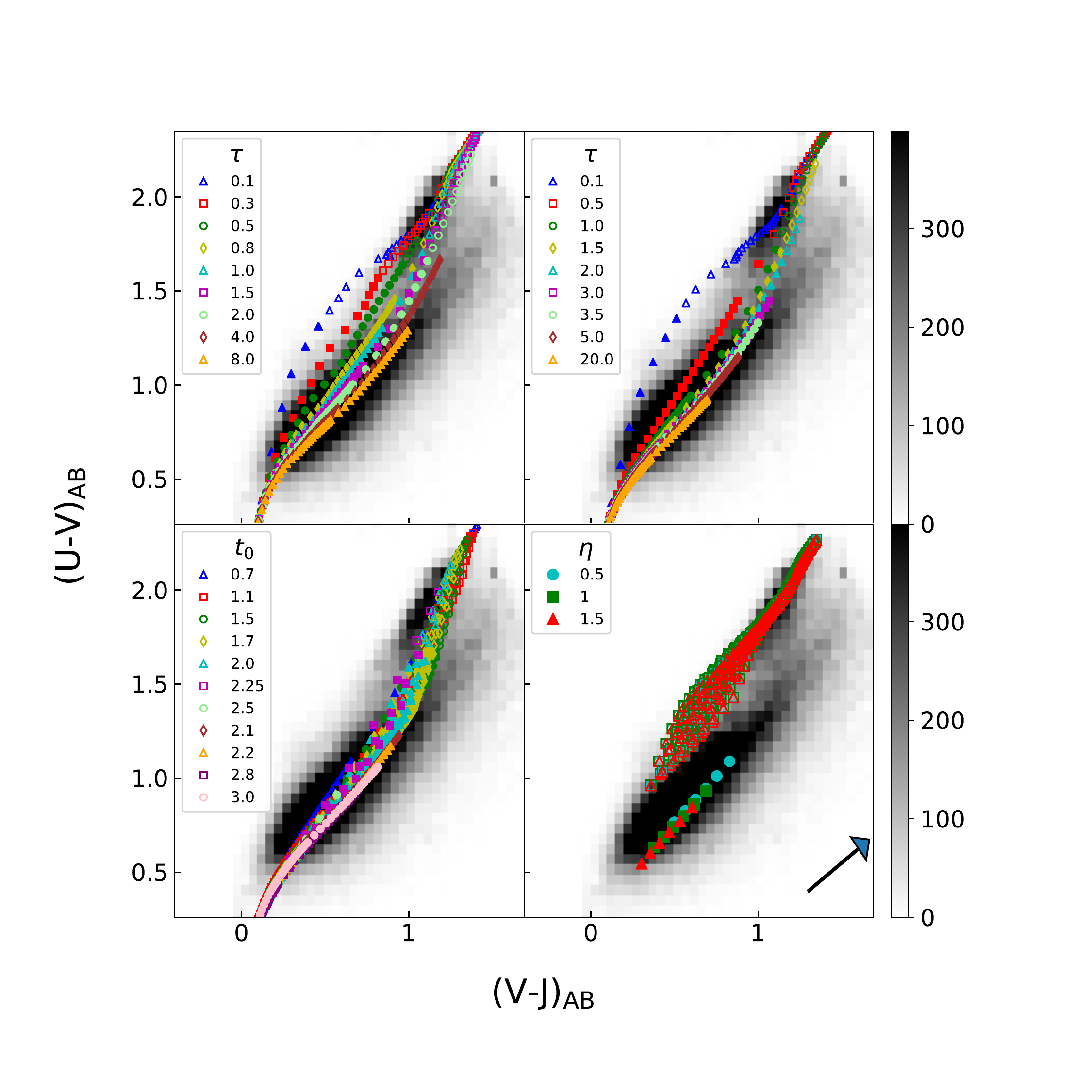}}
   \caption{Two-colour diagram for galaxy models, based on the BC03, solar metallicity SSPs, under different prescriptions for their SFH (coloured symbols), superposed on the distribution of rest-frame colours of galaxies of the SUDARE (COSMOS + CDFS) sample (grey scale). The four panels show the models computed with exponentially decreasing SFH (top left), 
delayed exponential SFH (top right), log-normal SFH (bottom left), and inflating + quenched models (bottom right). 
The colour and symbols encoding are labelled in each panel.
For the log-normal SFH, models with the same  $t_0$  but different value of the $\tau$ parameter are plotted with the same symbol (shape and colour). In all panels, filled symbols show models with sSFR $ \geq 10^{-11} \msun$/yr, eligible to be classified as star forming objects, while empty symbols show models with sSFR $< 10^{-11} \msun$/yr, which would be classified as passive galaxies.
The arrow in the bottom right panel shows the direction of the reddening vector. }
\label{fig_two_cols}
\end{figure*}

\subsection{Star formation history}

In the standard approach a galaxy is viewed as a collection of stellar populations of  different ages with the age distribution reflecting the SFH described by a parametric analytic relation, for example exponentially decreasing functions with different e-folding times. The SFH in each galaxy is characterized by two parameters: the age of its oldest stars and the parameter of the adopted analytic relation.
The packages currently used to derive age and age distributions in galaxies from their SEDs usually implement this kind of description. 

Both cosmological SFHs considered in this work were constructed to reproduce the observed relation between the SFR and stellar galaxy mass (galaxy main sequence), although under different conceptions. In the \cite{Peng_2010} model all galaxies evolve along this galaxy main sequence locus while they are star forming, until quenching sets in and they rapidly become  red, passive objects. Conversely, in the \cite{Gladders_2013} model galaxies spend most of their lifetime on the main sequence, but do not necessarily evolve along it. In addition, no abrupt quenching occurs; rather, at some point in the evolution, the SFR undergoes a gradual downturn, and the galaxies turn red while their SFR extinguishes. 

In order to test the sensitivity from the SFH of the constraints on the DTD, as derived from the correlation of the SNIa rate with the parent galaxy properties, we consider the following alternative SFH descriptions:

\vspace{0.5truecm}
\textit{Standard}:
\begin{itemize}
\item exponentially decreasing SFH:
 
$\psi(t) \propto exp (-t/\tau)$  with $\tau$ ranging from 0.1 to 8 Gyr;

\item delayed exponential \citep{gavazzi_2002}:

$\psi(t) = \frac {t}{\tau^2} \, exp \left( \frac{-t^2}{2 \, \tau^2} \right)$ 
 with $\tau$ ranging from 0.1 to 20 Gyr.
\end{itemize}

\textit{Cosmological}:
\begin{itemize}
\item log-normal SFH : 

$ \psi(\tu) \propto \frac{1}{\tu \, \tau} \cdot {\rm exp} (- \frac{(ln\,  \tu - t_0)^2}{2 \tau^2}) $ 
 
regulated by the two parameters $\tau$ and $t_0$, respectively, controlling the width of the distribution and the cosmic epoch at which the SFR peaks \citep{Gladders_2013};

\item inflating+quenched SFH, following \cite{renzini_2016} with two regimes:

$\psi(\tu) =  2.5 \times \eta\, M(\tu)\, (\frac{\tu}{3.5})^{-2.2}$

during the active phase, which lasts up to an abrupt quenching occurring at some cosmic epoch $\tu$, and is thereafter followed by pure passive evolution. In this equation $\eta$ is an adjustable parameter of the order of  unity, and $M(\tu)$ is the galaxy stellar mass at cosmic time $\tu$.
\end{itemize}

\noindent
In all these relations time variables are expressed in gigayear. 
We point out that in the standard descriptions the independent variable ($t$) is the time since the beginning of SF in a galaxy, which can occur at any epoch up to the current age of the Universe; however in the cosmological formulations the independent variable  ($\tu$) is the age of the Universe.  In the following we detail the implementation in our modelling of the cosmological SFH models.

\subsubsection{Log-normal SFH model}

In this alternative, the galaxy SFR grows to a maximum and then decreases, following a log-normal law. The cosmic epoch at which the maximum occurs, and the width of the log-normal relations are peculiar to individual galaxies. \cite{Gladders_2013} used the observed distributions of the specific SFRs of local galaxies, together with the cosmic SFR density
in the redshift range  $ z = 0 - 8$ to infer the distribution of the two parameters. We considered the results presented in Fig.~9 of Gladders et al., constrained by objects up to redshift 1. In this figure most galaxies occupy a triangular region in the ($\tau$ , $t_0$) space, which we sample with a number of discrete values. The variety of SFHs considered for this model  are shown in Fig. \ref{fig_abramo}. As $t_0$ increases the SF is more and more delayed, and at fixed $t_0$ several possible values of the parameter $\tau$ modulate the distribution of the stellar age. At each cosmic epoch,  galaxies with a variety of SFHs coexist, implying a wide distribution of galaxy colours. 

\subsubsection{Inflating+quenched SFH model}

The formulation for this model is taken from \cite{renzini_2016}, who has stated its applicability for $\tu \gtrsim 3.5$ Gyr. Presumably, galaxies form stars at earlier epochs, but at about 3.5 Gyr they are found on the galaxy main sequence, and from that time they evolve according to the given law until quenching occurs.  For galaxies forming stars with this SFH, at any cosmic time $\tu$, the fraction of mass in stars with  $ AGE = (\tu-t) $ is given by

\begin{equation}
f(AGE) = \frac {1}{M(\tu)} \frac {{\rm d} M} {{\rm d}AGE} = \frac {39.35\,\eta\, (\tu-AGE)^{-2.2}} {\rm {exp} [32.8\,\eta\,[(\tu-AGE)^{-1.2}-\tu^{-1.2}]}
,\end{equation}   

\noindent
which, at fixed $\eta$, only depends on the cosmic time $\tu$. In other words, in this model, all star forming galaxies at fixed redshift  have the same age distribution, and  their colours span a narrow range, reflecting only a spread in metallicity. Additional colour variance at fixed redshift could be attributed to very different contributions from the stellar populations formed at $\tu \lesssim 3.5$ Gyr; however, this is not consistent with the fact that in the inflating model most of the SF in star forming galaxies occurs at epochs later than 3.5 Gyr, when galaxies are seen on the main sequence.   Rather, we explore the effect of assuming different values of $\eta$, under the hypothesis that all galaxies follow the main sequence up to quenching, but with different {\it stamina}.  Figure \ref{fig_peng} shows the age distributions of the inflating SFH models at different cosmic epochs: we can see that when $\eta$ increases, at given cosmic time the galaxy hosts younger stellar populations. However, at late cosmic epochs even the most active star forming galaxies have a significant fraction of relatively old stars, since they have been sitting on the main sequence for a long time. 

\begin{figure}
\centering
\resizebox{\hsize}{!}{
\includegraphics[angle=0,clip=true]{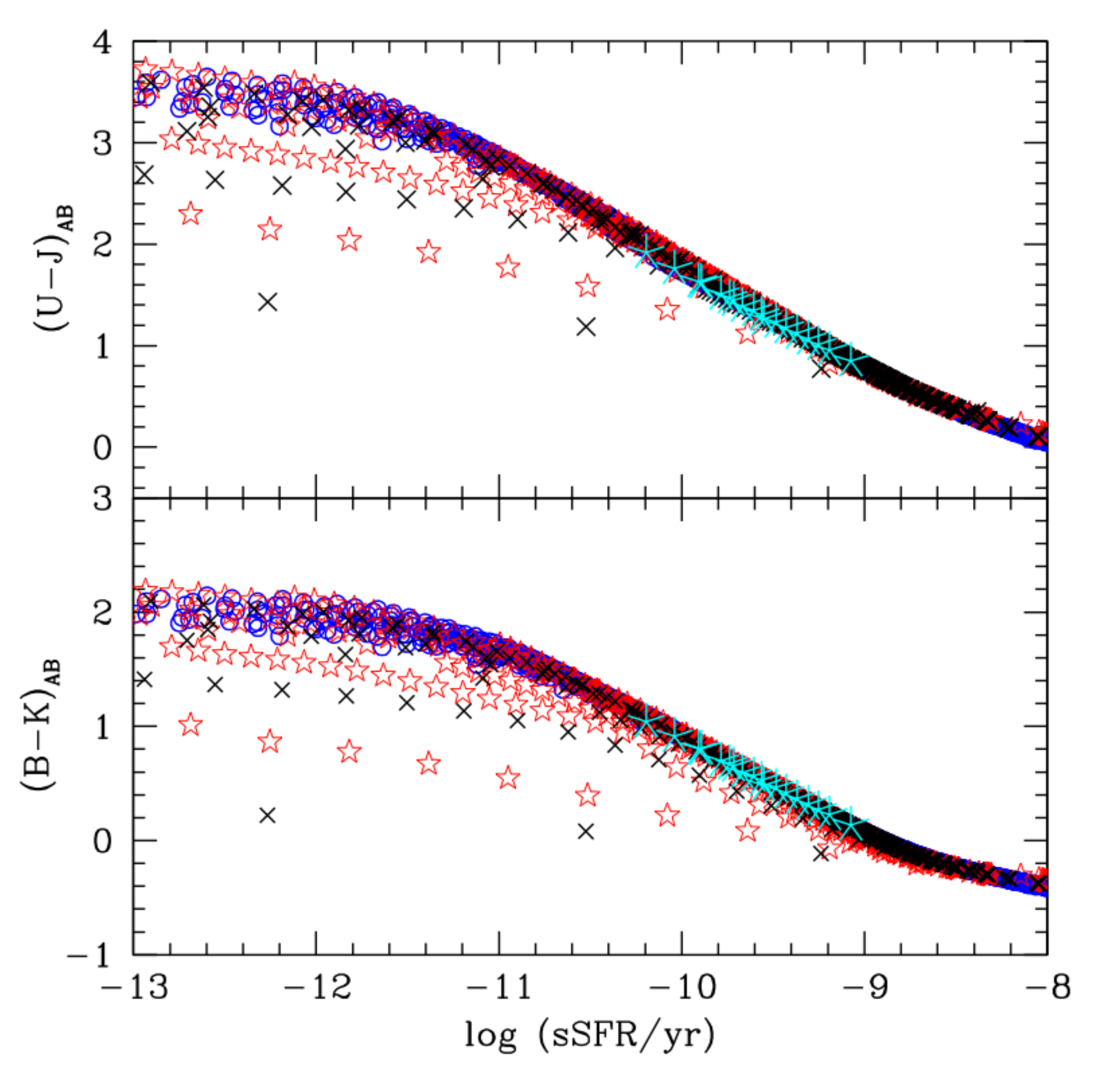}}
\caption{Integrated $U-J$ (top) and $B-K$ (bottom) colour as a function of the sSFR of model stellar populations with different SFHs:
log-normal (blue circles), exponentially declining (red stars), delayed exponential (black crosses), and inflating SFH (cyan asterisks).
There is a close correspondence between the integrated colour and the sSFR, which is largely independent of the SFH law. Solar metallicity BC03 SSP models  were used to compute the colours.} 
        \label{fig_ssfr_col}
        \end{figure}

\begin{figure*}
\centering
\resizebox{\hsize}{!}{
\includegraphics[angle=0,clip=true]{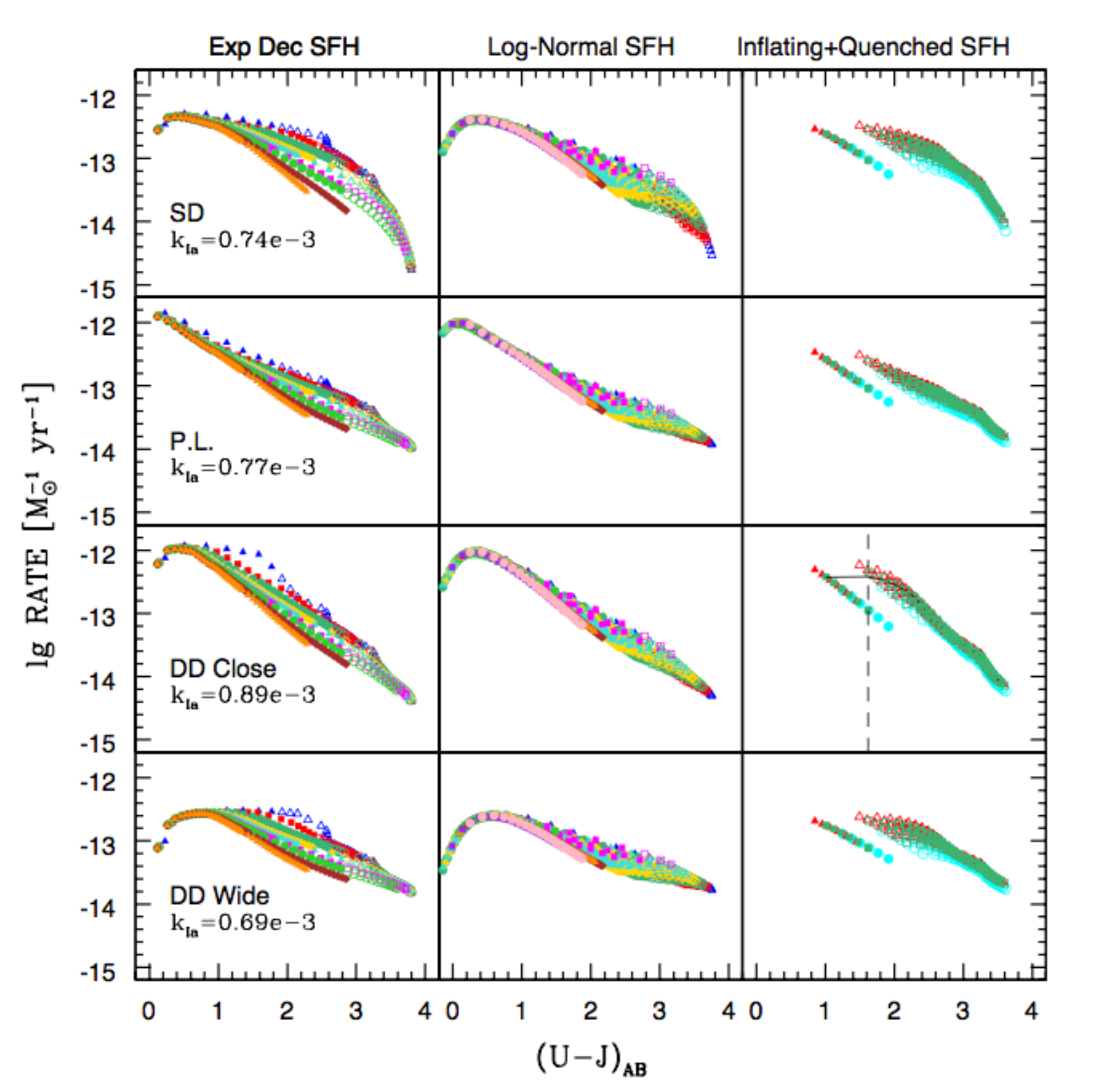}}
   \caption{Models for  the scaling of the SNIa rate per unit mass of formed stars with the $U-J$ colour of the parent galaxy, for different assumptions on the SFH and the DTD. Vertical panels share the SFH law as labelled on top, and horizontal panels share the adopted DTD, which are labelled in the left panels together with adopted values for \kia\ in $\msun^{-1}$. 
The symbols and colour encoding is the same as in Fig. \ref{fig_two_cols}, with star forming and passive models shown as filled and empty symbols respectively.  In the right panel relative to the DDC DTD, we show the evolutionary line of a model which undergoes quenching at $z=0.75$ when the universe was $\sim$ 7 Gyr old. In the same panel the vertical dashed line connects models with the same colour, but different SNIa rates (see text).}
             \label{fig_ria_umj}
\end{figure*}
                
\section{Integrated colours} 

In order to analyse the correlation between the SNIa rate and the colours of the parent galaxies we need to use models which well represent the spectrophotometric properties of the observed sample.
The SFH description gives the age distribution for a given galaxy but to obtain its colours we need to combine this information with the predictions for evolutionary models of simple stellar populations 
(SSP ).
The SSP models present their own systematic differences due to different input stellar tracks and bolometric corrections to temperature  transformations. A discussion of the related uncertainties is beyond the aims of this paper. However, we performed a basic check of the results obtained with three frequently used SSP models sets, namely: (i) the \cite{Bruzual:2003kx} library (hereafter BC03),
 specifically the models based on the Padova 1994 tracks coupled  with the STELIB spectra; 
(ii) the \cite{maraston+2010}  SSP models;   and (iii) the Padova SSP models based on the \cite{Marigo_2008} set of isochrones (hereafter MG08), as obtained with the CMD web tool\footnote{stev.oapd.inaf.it/cgi-bin/cmd}.
A brief report of this analysis is included in Appendix A and the results are summarized in Sect.~7.
All these sets model the properties of stellar populations of single stars. Spectrophotometric models for stellar populations which include binaries are available in the literature \citep[e.g.][]{Eldridge_2017}, but we do not use these models because they are bound to adopt additional parameters (e.g. binary fraction, distribution of separations, and mass ratios) which have some degree of uncertainty. In addition, we verified that, in combination with the SFHs adopted in this work, the \cite{Eldridge_2017} models do not reproduce well the colours of our sample galaxies (see Appendix A).

We found that the BC03 set with solar metallicity well represents the distribution of the galaxies in the $U-V$ versus $V-J$ rest-frame colour (cf. Fig.~\ref{fig_two_cols}) for all the different SFH models. Therefore, in the following we adopt BC03 as the reference SSP model set to calculate the galaxy integrated colour.

Inspection of Fig.~\ref{fig_two_cols} shows that these models nicely reproduce the general features of the observed galaxy colours by showing a correlation between the two colours which well matches the main axis of the observed distributions, and by reproducing the location on this plane of the star forming galaxies when the model sSFR is high and that of passive galaxies when the model sSFR is low. We point out that the galaxy colours in Fig.~\ref{fig_two_cols}  have not been corrected for reddening. The direction of the reddening vector, derived from the \cite{cardelli_1989} extinction curve, is shown as a blue arrow in the bottom right panel, and a similar direction applies for the \cite{calzetti_2001} extinction law. On this plane, the effect of both the Milky Way and of the internal absorption is that of shifting the galaxies to the red along the main axis of their distribution. As is well known, reddening is likely responsible for the red extension of the blue galaxy sequence, but it hardly produces a scatter around the main locus. On the other hand such scatter can be ascribed to some differences in the metallicity, given the high sensitivity of these colours to Z  (cf. Fig.~\ref{fig_cc_z}). 
 
Although the main features of the distribution of the models on this plane are similar in the four panels, some differences can be noticed. Some models computed with the standard SFH laws (top panels) fall in a region of red  $U-V$ for $V-J<1$  where observed galaxies are scarce (e.g. the blue triangles). These models have short timescales $\tau$, so that their colours evolve fast, taking only $\lesssim$ 3  Gyrs to reach the region of passive galaxies. Therefore, if  galaxies with short SF timescales formed at high redshift, by z $\simeq 1$ they will already have reached the passive region of the two-colour diagram.
In other words, it is possible to reconcile the colour distribution of the galaxy population with the standard SFH laws adopting an adequate distribution of the (AGE,$\tau$) parameters. 
Similarly, the quenched models in the bottom right panel of Fig.~\ref{fig_two_cols} at  $V-J<1$ fall in an underpopulated region.  But it only takes $\lesssim$ 1 Gyr to all quenched models to reach the region of passive galaxies, and therefore the transition region between the blue and red galaxies indeed is expected to be underpopulated. The log-normal SFH models (bottom left) reproduce very well the high density regions of the observational plane, with a good match of both star forming and passive galaxies. For this class of models the distribution of stellar ages in individual galaxies is  wide, so that on this plane they behave like the standard models with long timescales.

In summary, all the SFH laws considered in this work appear consistent with the distribution of galaxies on the two-colour plane, and  a more extended and punctual analysis of the galaxies distribution is required to obtain firm
constraints on the SFH in galaxies. This is beyond our scope; we rather focus on the ability of our models to reproduce the general features of the colours of the sample galaxies. 

For each of the SFH formulations considered in this paper, Fig. \ref{fig_ssfr_col} shows that there is a nice correlation between integrated colours and the sSFR. Furthermore, the different SFH laws remarkably describe the same correlation between sSFR and colour, except for the exponentially declining and the delayed exponential SFH with short timescales. For these latter models,  the decline of the sSFR as time increases is fast compared to the growth rate of the colour. On the other hand, Fig. \ref{fig_two_cols} shows that the exponential SFH models with short timescale fall in an underpopulated region of the two-colour diagram, unless their $(U-J) \gtrsim 2.7$. In other words, this SFH description could be adequate for passive galaxies with very red colours and very low sSFR. We conclude that using the $U-J$ age tracer is equivalent to using $B-K$, and largely equivalent to using the sSFR parameter.
The predicted correlation is used in Sect. 7 to infer sSFR from colour of local galaxies. 

\section{SNIa rate versus galaxy colours}

In this section we illustrate the model predictions for the correlation between the SNIa rate and the colours of the parent stellar population, used as a diagnostic tool for the DTD.  
Fig.~\ref{fig_ria_umj} shows the model correlations of the SNIa rate with the $U-J$ colour of the parent galaxy for various formulations of the SFH and the four DTD options shown in Fig.~\ref{fig_dtds}. Delayed exponential SFHs, not shown in this figure, provide trends very similar to the exponentially declining models. Clearly both the DTD and the SFH bear upon the resulting trend, since both the rate and the colour depend on the age distribution of the stellar population, but have different sensitivities. Thus, galaxies with different SFHs may have the same $U-J$ colour, but different SNIa rate. In other words, the SNIa rate and the colour trace the average age of the parent stellar population, but they do so in different ways. 

Irrespective of the SFH, some features of this correlation are peculiar to the DTD: the single degenerate channel exhibits a fast drop at red colours, for old passive galaxies, where only events with long delay time take place. We recall that in the SD model, such events are produced in systems with a high mass primary because of the need to reach the Chandrasekhar limit by accreting the envelope of the low mass secondary, which fills the Roche Lobe at late delays.  
The P.L. and the DDC DTDs produce a relatively narrow correlation, steeper for the DDC model with respect to the P.L. model. The DDW DTD produces a  relatively flat correlation, with a drop of only a factor of $\sim$ 10 between the bluest and the reddest galaxies.

Some features of the correlation between the SNIa rate and the $U-J$ colour of the parent galaxy are instead peculiar to the SFH: exponentially decreasing laws with different e-folding times produce a relatively large variance of the rate at fixed $U-J$ colour, especially at $1.5 \lesssim (U-J) \lesssim 2.5$, which is typical of Spiral galaxies. Compared to the standard SFH models, the log-normal laws result in a tighter correlation; models with very different widths and peak times describe almost the same scaling of the SNIa rate with the $U-J$ colour of the parent galaxy.

The inflating+quenched SFH models predict a very distinctive behaviour, with two separate sequences, one for star forming galaxies (at bluer colours) and one for passive galaxies. In this class of models, galaxies evolve along the star forming sequence until quenching occurs, at which point they rapidly become red. 
In Fig. \ref{fig_ria_umj} the models adopting the inflating+quenched SFH are plotted for $t_U \gtrsim 7$ Gyr. In a galaxy sample which includes objects up to high redshift there should appear galaxies with the same colour, but lower SNIa rates if star forming, larger SNIa rates if passive, irrespective of the DTD.
For illustration, the panel relative to the DDC model shows an evolutionary line for galaxies which underwent quenching when the universe was 7 Gyr old. The gap between the two sequences in the right panels of Fig. \ref{fig_ria_umj} corresponds to a 0.1 Gyr time elapsed from quenching, a process assumed instantaneous. A less abrupt quenching would produce a more gentle evolution from the blue to red sequence. However, it is expected that because of the large contribution of events at delay times of  a few hundred million years, the drop of the SNIa rate lags behind the colour evolution, and there should be some  relatively red galaxies with a high SNIa rate. In a galaxy sample which includes high statistics for objects up to redshift $\sim$ 1 (when the universe was $\sim$ 6 Gyr old) we may detect these two sequences, with galaxies with the same $U-J$ colour showing a higher SNIa rate if quenched, compared to  galaxies which are still star forming. The size of the effect is illustrated in the right panel of Fig. \ref{fig_ria_umj} of the DDC delay time distribution, where a vertical line  connects two model galaxies with the same colour, but different SFH: one star forming in the local universe and one quenched at $z =0.75$ in the very early post-quenched life. The two galaxies happen to have the same colour ($U-J$ = 1.6), while the SNIa rate in the quenched galaxy is higher  because its average age  ($<AGE> \simeq 1.1$ Gyr) is younger than that of the local still star forming galaxy ($<AGE> \simeq$ 3.9 Gyr). 

\begin{figure*}
\centering
\resizebox{\hsize}{!}{
\includegraphics[angle=0,clip=true]{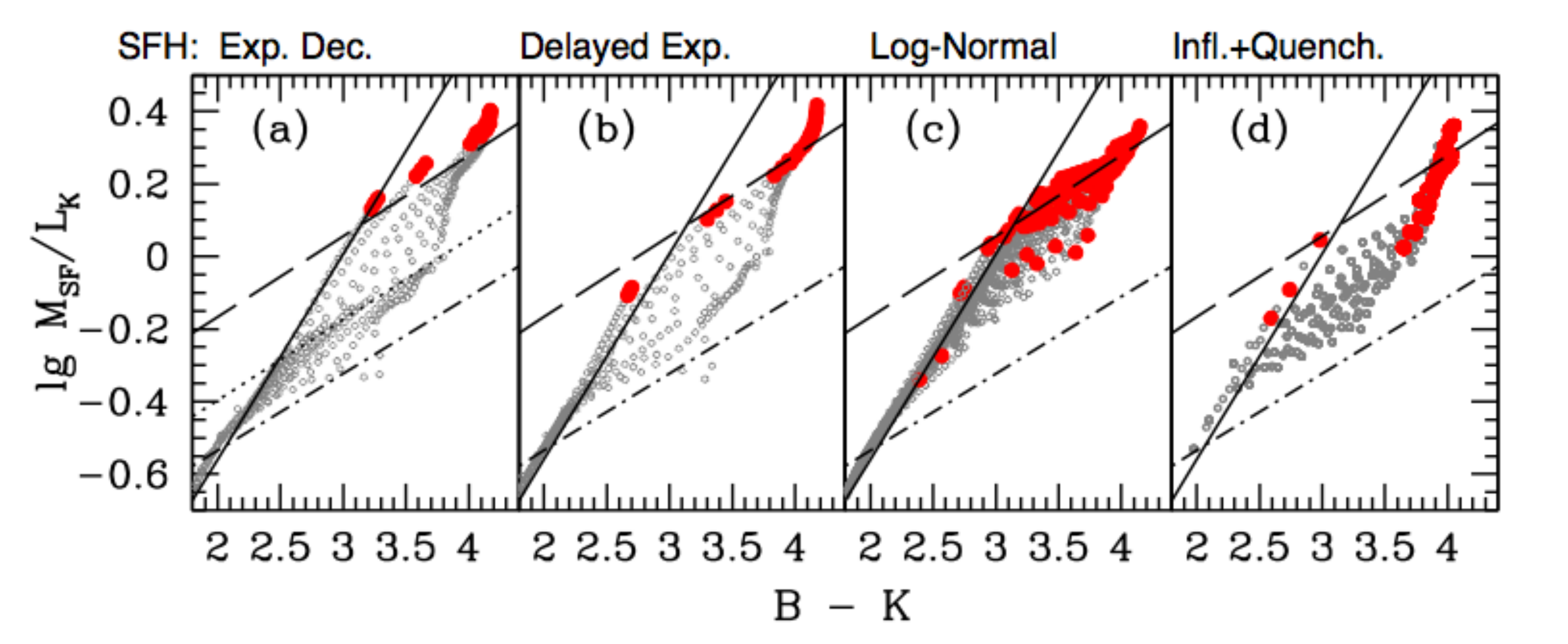}}
   \caption{Model predictions for the ratio of the mass of formed stars to the $K$-band luminosity (both in solar units) as a function of the $B-K$ colour (in VEGAMAG) of the parent stellar population for different SFH laws, as labelled on top.
An absolute magnitude of $M_{K,\odot}=3.26$ has been used to compute the mass-to-light ratio of the stellar populations.
Filled red circles highlight models with $AGE \geq 12$ Gyr in panels (a) and (b), and models with $t_{\rm U} \geq 13$ Gyr in panels (c) and (d).  The lines show various options to derive galaxy stellar masses from the $B-K$ colour and $K$-band luminosity: the dotted line in panel (a) shows the BJ01 relation for a Salpeter initial mass function (IMF); in all panels the dash-dotted line shows the BJ01-a regression, while the solid and the dashed line are combined in the piece-wise relation. See text for more details.}
             \label{fig_molk}
\end{figure*}

\begin{figure}
\centering
\resizebox{\hsize}{!}{
\includegraphics[angle=0,clip=true]{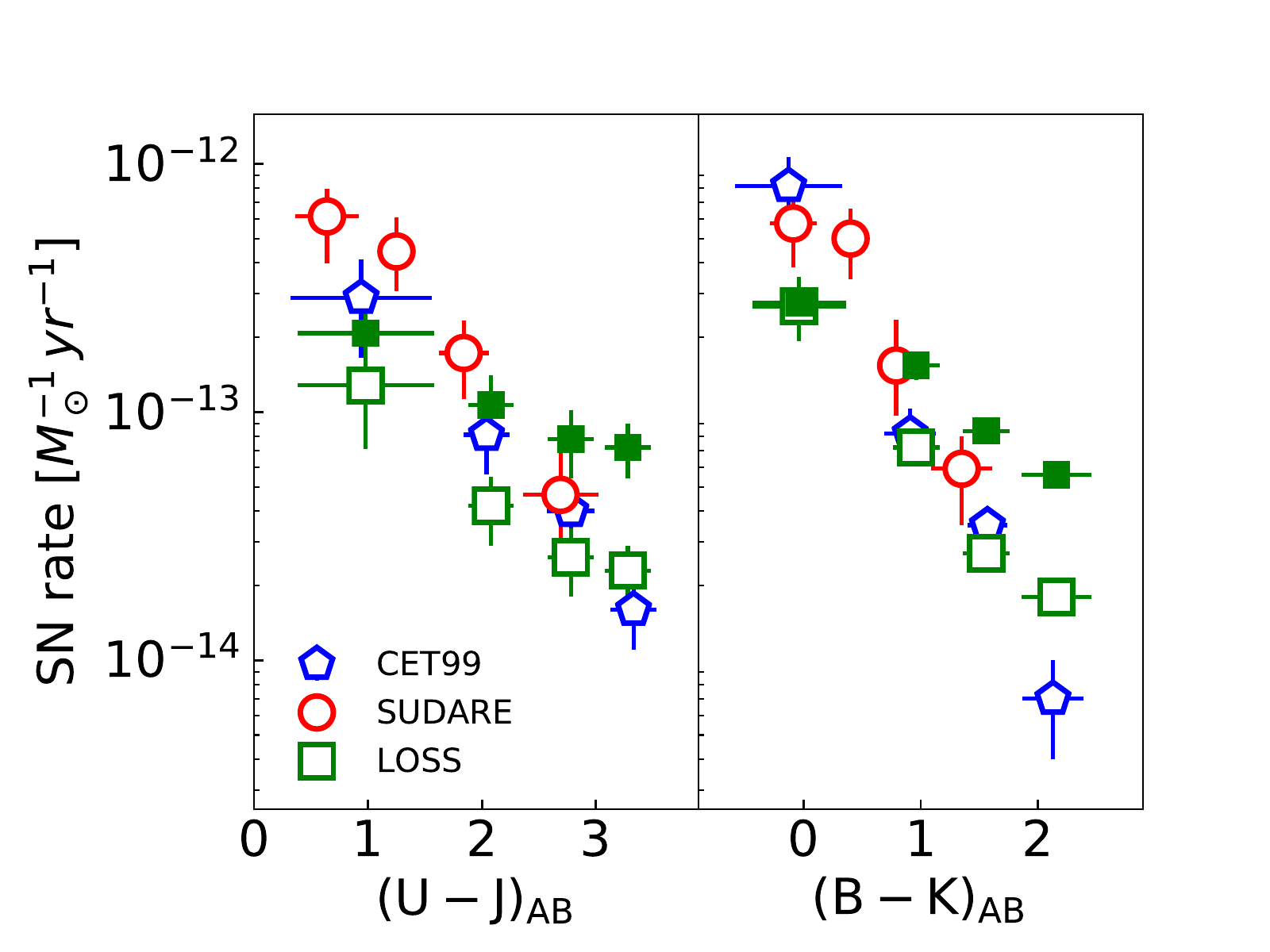}}
\caption{Observed correlations between the SNIa rate and the parent galaxy $U-J$ (left) and $B-K$ (right) colour in the SUDARE, CET99, and LOSS surveys. For the local surveys the stellar masses of the galaxies have been evaluated with the piece-wise relation discussed in the text. For the LOSS survey we also show as filled symbols the rate normalized to the mass derived from the BJ01-a relation.
Colours are shown in the AB magnitude system. The horizontal error bars show the 1$\sigma$ width of the galaxy distribution within each colour bin; the vertical error bars show the uncertainty on the rate from the statistics of the events. }
        \label{fig_data}
        \end{figure}

\section{Observed rates and mass-to-light ratios.}

The observational counterpart of the correlations plotted on Fig.~\ref{fig_ria_umj} are the rates per unit mass in galaxies of different  colours. 
To derive constraints on the DTD, it is necessary to collect a big sample of galaxies to provide a significant number of SNIa events within each colour bin. For each galaxy, the rest-frame, dereddened colour has to be determined along with the stellar mass. 

The mass of a stellar population evolves with time because of the progress of SF and to the mass return from stellar winds and supernovae. Rather than the current stellar mass, it is more appropriate to consider the total mass of formed stars ($M_{\rm SF}$), that is the integral of the  SFR over the total galaxy lifetime, writing Eq. (\ref{eq_rate}) as

\begin{equation}
\frac{\dot{n}_{\rm Ia}(t)}{M_{\rm SF}} = \kia \frac {\int_0^t {\psi(t-\td)\,\fia (\td)\,\rm{d} \td}}{\int_0^t{\psi(t)\,\rm{d} t}}.
\label{eq_rate_nor}
\end{equation}

\noindent
When  normalizing the rate to the current mass of the stellar population, the mass reduction should be taken into account, a factor which is a function of time,  SFH,  IMF, and the initial-final mass relation of single stars \citep[see][]{greggio_book}.
Thus the scaling between the DTD and the observed rate would be less straightforward, and more prone to the choice of the input ingredients of the modelling. Some modelling is also required to apply Eq. (\ref{eq_rate_nor}), 
since the integral of the SFR  is not directly measured and rather derived from the galaxy luminosity. This is a critical issue as we show in the following.

In this paper we consider SNIa rates measured on different searches and galaxy samples. Galaxies in the SUDARE sample were analysed using the  EAZY code to determine the redshift, and the FAST code \citep{Kriek2009} to determine the dereddened colours and the mass of formed stars by fitting their SED from the UV to the IR.  In general, the best fit SFH and then its integral (\msf) depends on the specific SFH description adopted by FAST to fit the SED. In practice, we verified that in the SUDARE sample the observational correlation remains the same when using the exponentially decreasing or the delayed  SFH law. 

The photometric data available for the local samples  (LOSS and CET99) are too sparse to perform the SED fitting and derive the SFH of all galaxies. Therefore, to estimate the stellar mass of each galaxy we adopt a relation between the $M/L_{K}$ ratio and the  $B-K$ colour of the parent stellar population, as often done in the literature \citep[e.g.][]{Mannucci_2005,Li_2011}. However, the calibration of this relation requires some discussion. 

Fig.~\ref{fig_molk} shows the ratios of the mass of formed stars to the $K$-band luminosity of model galaxies computed with  a Salpeter IMF (from 0.1 to 100 \msun) as a function of their $B-K$ colour. The four panels refer to the different kinds of SFH considered here. Since we  use these models to derive the stellar mass of galaxies in the local universe, we highlight with red filled circles those with old ages, i.e. models with $AGE \geq 12 $ Gyr for the standard SFHs (panels (a) and (b)), and models with $t_{\rm U} \geq 13$ Gyr for the cosmological SFH laws (panels (c) and (d)).
In the  standard models the $AGE$ parameter is the time elapsed since the beginning of SF, and even in the local universe this could be short for galaxies of the latest types.
Nevertheless, in panels (a) and (b) we highlight the models with old age for a direct comparison to the Bell and De Jong (2001, hereafter BJ01) 
relation, which has been often adopted in the literature to derive the stellar mass from the galaxy luminosity.

The dotted line in panel (a) of Fig.~\ref{fig_molk} shows the BJ01 regression for exponentially decreasing SFH models with different e-folding times, all at an age of 12 Gyr, and assuming the same IMF and set of SSP models as in our computations (coefficients are taken from Table 4 in BJ01). Its slope well matches our 12 Gyr old models in panel (a), but the Bell and De Jong masses are systematically lower than our masses by a factor of $\sim 0.6$  (see dashed line in Fig.~\ref{fig_molk}). This shift is largely a consequence of a different definition of the stellar mass, while we consider \msf , BJ01 refers to the current mass of the stellar population, i.e. taking into account the mass recycling during the evolution of the galaxy. The mass reduction factor of a SSP with a Salpeter IMF at an age of 12 Gyr is $\simeq 0.7$ \citep{greggio_book}; the small residual discrepancy likely results from other different choices in the ingredients of the computations. Nevertheless, we stress that the slope of the relation between the mass-to-light ratio and the integrated colour of the stellar population is the same in ours and in the BJ01 models, when referring to the same SFH and age. 
The dash-dotted lines in Fig.~\ref{fig_molk} show the regression from BJ01 used in \cite{Mannucci_2005} to convert $K$-band luminosity into stellar mass (hereafter BJ01-a). This is the preferred regression in BJ01 for 12 Gyr old galaxies. This regression is based on a somewhat different assumption on the SFH of the model galaxies, but most importantly it assumes a modified Salpeter IMF, flattened at the low mass end.  The slope of this latter regression is virtually the same as for the exponentially declining models, but at given luminosity it implies a stellar mass that is lower by a factor 0.7 (0.15 dex) because of the different IMF.

When analysing the correlation between the SNIa rate and the colour of the parent galaxy to constrain the DTD, a zero point shift of the mass-to-light ratio results in a different value for \kia, but has no effect on the shape of the DTD.  On the other hand, changing the slope of the regression may have important consequences.
Actually, in the local universe there are galaxies whose blue colour demand young ages, represented as grey circles in the four panels of Fig.~\ref{fig_molk}. In fact in the RC3 catalogue almost half of the galaxies are bluer that $B-K$ = 3;  a fair number of objects are as blue as $B-K \sim 2$. For these blue galaxies, the log-normal SFH models cluster along a very steep relation, which is also consistent with the inflating SFH models, during the active phase. The quenched models follow a parallel relation, shifted to redder colours; they describe a very different trend compared to the old and red models resulting from the other SFH laws.  

In general, Fig.~\ref{fig_molk} shows that the $B-K$ colour may not be a good tracer of the mass-to-light ratio in local galaxies in contrast to what is usually assumed; this applies to standard SFH laws when allowing for an age spread and to the models constructed with the cosmological SFH. Nonetheless,
with the aim to highlight the impact of the adopted mass-luminosity relation, we construct a piece-wise relation tailored to the log-normal SFH models with $t_{\rm U} \geq 13$ Gyr (red dots in panel (c) of Fig.~\ref{fig_molk}); we use the combination of the solid (for blue galaxies) and dashed (for the red galaxies) lines shown in all panels. 
Although this representation is not spotless for the red galaxies, it improves the estimate of the mass-to-light ratio for the blue galaxies with respect to the BJ01 relation. We checked that for the SUDARE galaxies we find an overall consistency between the mass of formed stars obtained from the SED fitting and from the use of this piece-wise relation.

Fig.~\ref{fig_data} shows the measured rates versus the galaxy colours for the three independent surveys mentioned above. For the LOSS survey we show the effect of adopting different relations to evaluate the mass-to-light ratio at given $B-K$ colour:  filled symbols assume the  BJ01-a relation while empty symbols are derived with the piece-wise relation.
It turns out that the slope of the observed correlation is very sensitive to the adopted scaling of the mass-to-light ratio with the galaxy colour. For the blue galaxies, the mass-to-light ratios from the BJ01-a and from the piece-wise regressions are similar (see Fig.~\ref{fig_molk}), and so are the derived rates per unit mass in the bluest bin. As the colour becomes redder, the masses derived at fixed ($B-K,L_{K}$) from the BJ01-a relation are systematically lower than those estimated with the piece-wise relation and the rates per unit mass are evaluated to be higher, an effect which becomes more prominent as the $B-K$ colour gets redder.  
This systematic trend is also found for the correlation of the rate with the $U-J$ colour because red galaxies in $U-J$ are also red in $B-K$. 

For the two local surveys, the rates per unit mass as  functions of $U-J$ are consistent within the statistical errors when the same mass-to-light ratio regression is used; the same is true for the rates as a functions of $B-K$ but for galaxies in the bluest and the reddest bins, and the CET99 values point at a much steeper correlation with respect to LOSS. This difference could be related to the poor statistics of the CET99 sample, which counts only 10 and 6 events in the bluest and in the reddest bins, to be compared to the 14 and 73 events in the LOSS survey in the same colour bins. Good statistics of events is clearly the first requirement to derive useful information from these correlations. 

As mentioned above, the galaxy masses in the SUDARE survey are derived from the multicolour SED fitting, and are not prone to the ambiguity affecting the mass-to-light ratio as traced by the $B-K$ colour. Actually the SED fitting is designed to determine the age distribution, yielding a SFH for each galaxy. On the other hand, other systematic effects may introduce an uncertainty in the derived masses, e.g. errors in the estimate of the redshift and of extinction correction, but also some degeneracy in disentangling the contribution of the various stellar population components. In addition,  for the SUDARE galaxy SED fitting we use the FAST program with a  standard SFH description, and we cannot check whether using, for example a log-normal description of the SFH, would lead to a different estimate of the galaxy mass. Yet, as we can see in Fig.~\ref{fig_data} the rates from the SUDARE survey appear largely consistent to those of the local surveys. The agreement is very good when binning the galaxies in $B-K$ colour, while when binning in $U-J$ the SUDARE rates tend to be systematically higher than those measured in the local surveys. Given statistical and systematic uncertainties, we conclude that overall there is no tension among the results from the three surveys.

\section{Examining the diagnostic from the observed correlation}

As can be appreciated from Fig.~\ref{fig_ria_umj}, the unambiguous interpretation of the correlation requires knowledge of the SFH. The data in hand do not allow us to discriminate among the various SFH models for the galaxies in our data sample. However in this section we proceed by comparing models to observations to illustrate the diagnostic capabilities of this correlation once the SFH variable is fixed. 
To do this we consider the predictions of the log-normal laws in combination with the BC03, solar metallicity SSP set, which appears to well represent the SUDARE galaxy sample (see Fig.~\ref{fig_two_cols}).
We remark that we do not mean to favour any specific SFH description or SSP set of models. To assess the scenario which best describes the photometric properties of galaxies a much more thorough analysis than carried out in this work would be necessary.  

\begin{figure}
\centering
\resizebox{\hsize}{!}{
\includegraphics[angle=0,clip=true]{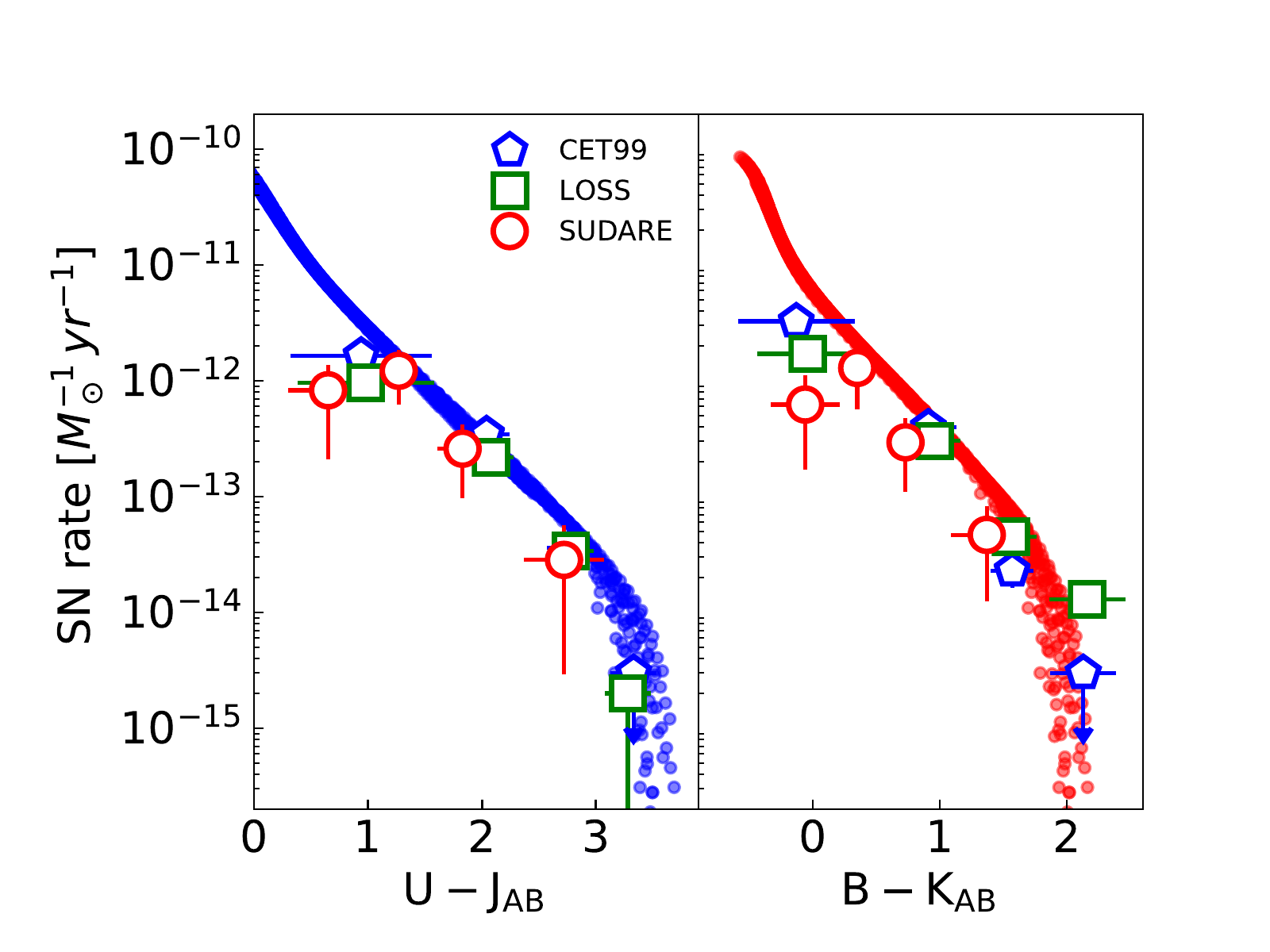}}
\caption{Rate of CC supernovae as a function of the colour of the parent stellar population for the log-normal SFH laws (filled dots) compared to the rates measured in the three independents surveys SUDARE (empty red circles), LOSS (green squares), and CET99 (blue pentagons). Error bars have the same meaning as in Fig. \ref{fig_data}.
The theoretical rates are obtained as the plain product of the specific SFR and $k_{\rm CC} = 0.005 \msun^{-1}$.}
\label{fig_rcc_color}
        \end{figure}

As a consistency check, Fig.~\ref{fig_rcc_color} shows the comparison between models constructed with the log-normal SFH and the rate of core collapse (CC) events measured in the three galaxy samples. The theoretical rates are computed by multiplying the specific SFR of our selection of log-normal SFH models by the number of CC progenitors per unit mass of the parent stellar population (\kcc). For a Salpeter IMF and progenitors with mass between 10 and 40 \msun, $\kcc=0.005 \msun^{-1}$; for this value there is a very good agreement between the models and the data.  Assuming a lower value of the minimum mass of the CC supernovae progenitor \citep[see e.g.][]{smartt:2009mq}, \kcc\ would be higher and the observed rates would be lower than the expectations. This may suggest that a fraction of the events go undetected, or equivalently, that the detection efficiency of the search is overestimated. The trend of the model rates with the colour of the parent galaxy fits the observations very well, therefore this effect should be approximately the same in blue and red galaxies. A relatively low value of \kcc\ was found in paper II for the SUDARE sample and is also found  for the local surveys in this work. On the other hand, the volumetric rates presented in paper I from the SUDARE sample support a mass range  between 8 and 40 \msun\ for the CC supernovae progenitors (\kcc=0.007) in combination with the cosmic SFR of \cite{madau:2014uf}. We do not have a ready explanation for this discrepancy, which could be related to the intrinsic uncertainty in the measurement of the galaxy masses, on the one hand, and of the cosmic SFR, on the other.  

Fig.~\ref{fig_rcc_color} shows that the log-normal SFH models adequately describe the scaling of the specific SFR with the colour of the parent galaxy. This also applies to the other formulations of SFH considered in this work, which is not surprising given the tight relation between the specific SFR and the colour of the parent galaxy shown in Fig.~\ref{fig_ssfr_col}, followed by all the SFH laws considered in this paper. 

\begin{figure}
\centering
\resizebox{\hsize}{!}{
\includegraphics[angle=0,clip=true]{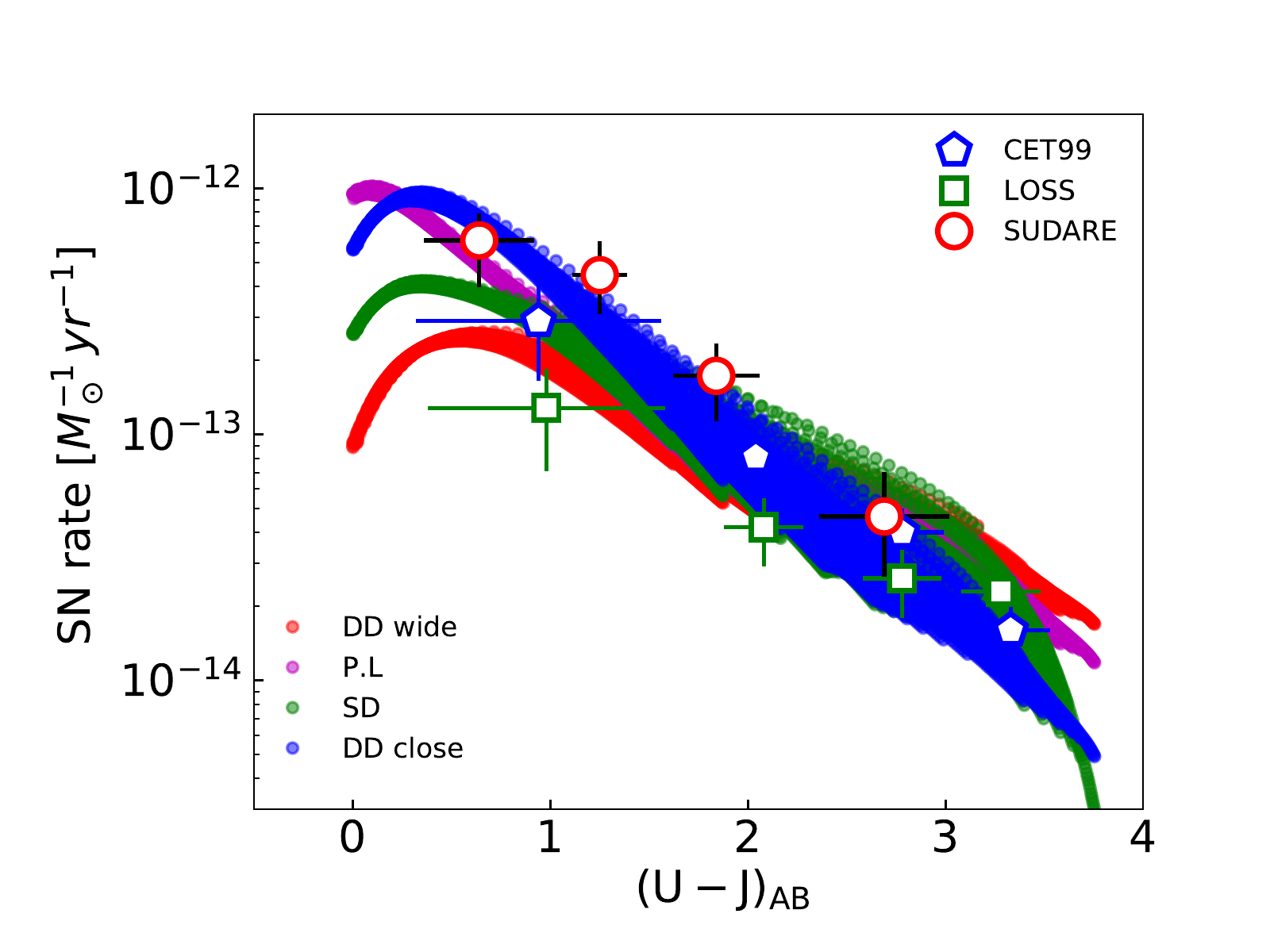}}
\caption{Models computed with the log-normal SFH  and the four options for the DTD considered here (filled symbols) compared to data (empty symbols). Predictions obtained with different DTDs are plotted with different colours as labelled.
For each DTD we adopt the best fitting value for \kia, namely  \kia=(0.74,0.77,0.89 and 0.69)$\times 10^{-3}$ 1/\msun\ for the SD, P.L., DD CLOSE and DD WIDE DTDs, respectively (see Appendix B). The symbols encoding of the observed rates is the same as in Fig. \ref {fig_data}. }\label{fig_corr_umj}
        \end{figure}

\begin{figure}
\centering
\resizebox{\hsize}{!}{
\includegraphics[angle=0,clip=true]{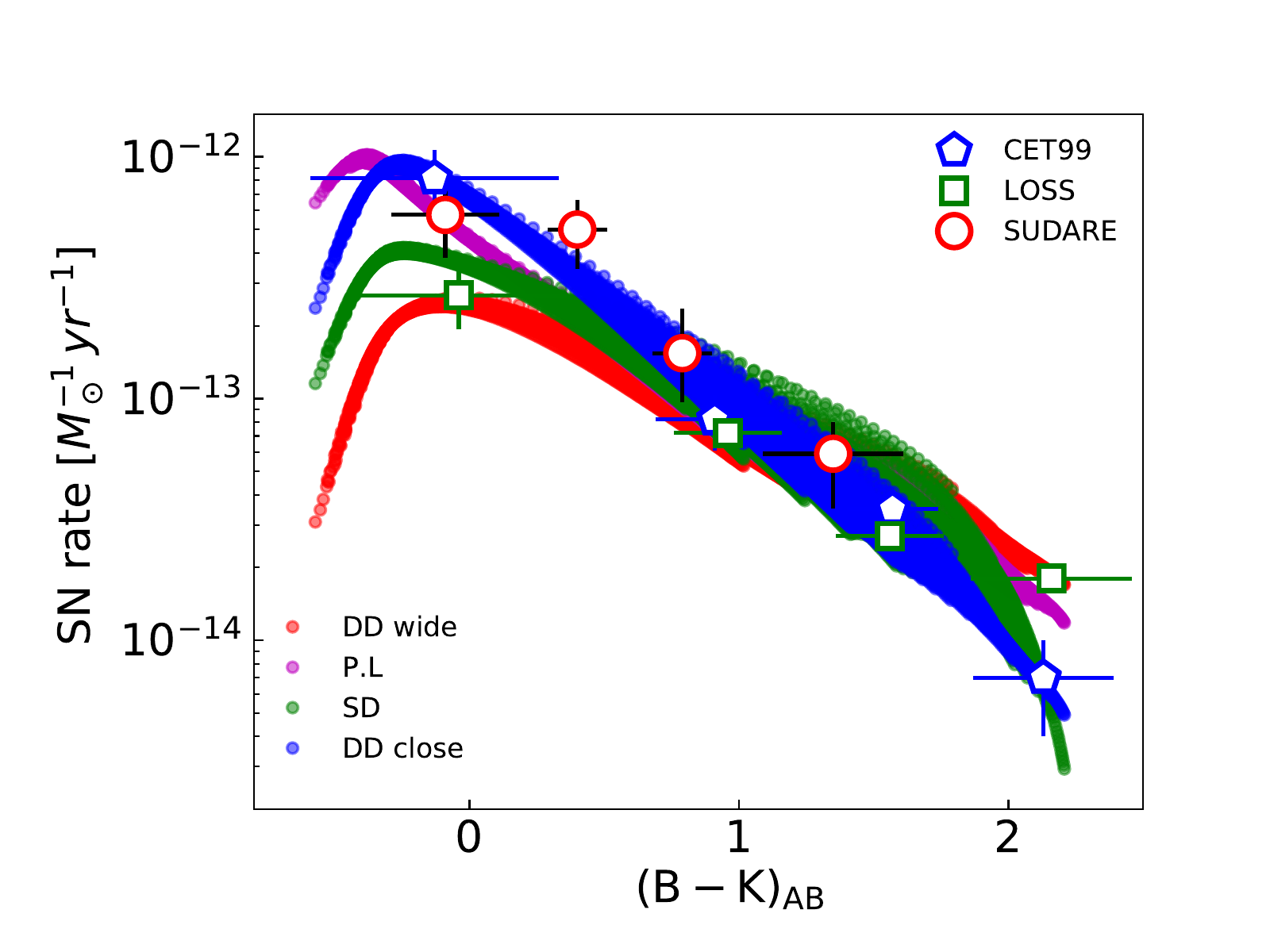}}
\caption{Same as Fig. \ref{fig_corr_umj} but for the galaxies binned according to their $B-K$ colour.}
        \label{fig_corr_bmk}
\end{figure}

\begin{figure}
\centering
\resizebox{\hsize}{!}{
\includegraphics[angle=0,clip=true]{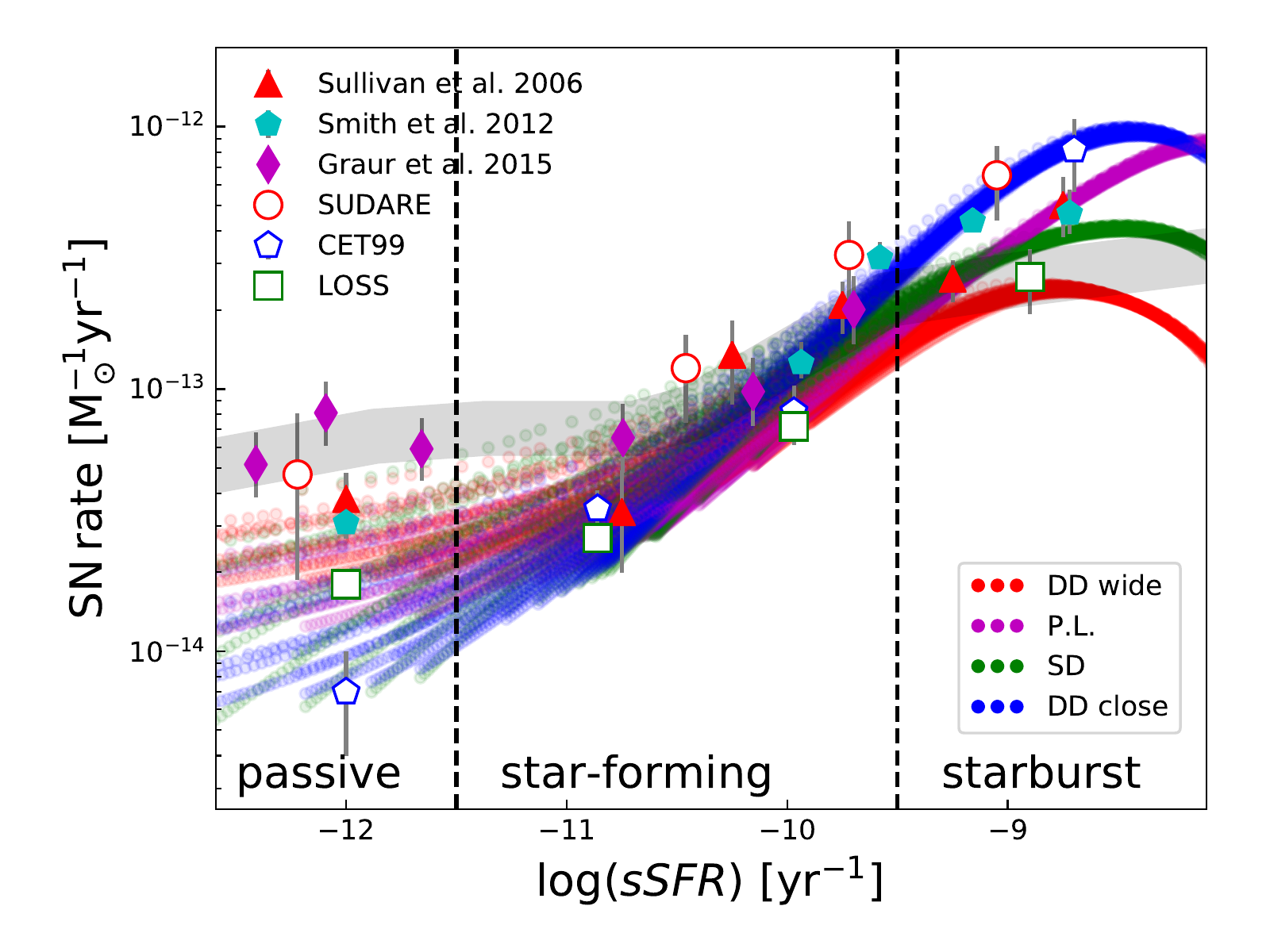}}
\caption{Correlation between the rate of SNIa and the specific SFR of the parent galaxy. Filled circles show our models computed with the log-normal SFH; the colour encoding is  the same as in Fig. \ref{fig_corr_umj}. Open symbols show our measurements on the SUDARE, CET99, and LOSS data. For the local surveys we adopt the piece-wise relation between the M/L$_K$ ratio and the $B-K$ colour. The filled triangles, pentagons, and diamonds show literature data (see legend in the top left corner). The grey stripe shows the result of the simulations in \cite{graur+2015}.}

        \label{fig_ria_ssfr}
\end{figure}

Figures \ref{fig_corr_umj} and \ref{fig_corr_bmk} show the comparison between the models and the measurements of the SNIa rate versus the $U-J$ and versus the $B-K$ colour for the three independent surveys.  The observed rates for the local samples are normalized to the stellar mass derived using the piece-wise relation.  The model rates are obtained using Eq. (\ref{eq_rate_nor}) and adopting a value of  \kia\ which fits the observed rates in galaxies with intermediate colour. Indeed, in galaxies with a flat age distribution ($\psi$ = const.)  the SNIa rate at late epochs is equal to 
the ratio of \kia\ to the age of the galaxy, since the DTD is normalized to unity  \citep[see][]{greggio_2005}.
The colours of such stellar populations are 
$(U-J)_{\rm AB}  \simeq 2 $ and  $(B-K)_{\rm AB} \simeq 1$. In Appendix B we describe in  detail the procedure adopted to derive \kia;  we only remark that the productivity determined in this way turns out of $\sim 0.8 \times 10^{-3} \msun^{-1}$  and that small variations result from different DTD models. In an SSP of 100 \msun, there are three stars with mass between 2.5 and 8 \msun\  for a Salpeter IMF; this implies that  the probability that one such star ends up as a SNIa should be of $\sim$ 3 \% to account for the observed level of the SNIa rate in galaxies.

This estimate is in excellent agreement with the result in Paper I from the analysis of the cosmic SNIa rate, while it is $\sim 30$\% lower than that derived in Paper II, where we found $\kia \sim 1.2 \times 10^{-3} \msun^{-1}$. We ascribe this discrepancy to the 
different procedure employed: in paper II we only considered the correlation in the SUDARE data with the $U-J$ colour, while in this paper we average over the three surveys and further average the results from the correlation with both  $U-J$ and $B-K$. 

Once the productivity is scaled to galaxies with intermediate colours, the bluest and reddest bins constrain the shape of the DTD. 
Inspection of Figures~\ref{fig_corr_umj} and \ref{fig_corr_bmk} shows that the available data do not allow us to draw a reliable conclusions on the DTD. The LOSS survey tends to favour a flat shape, while the CET99 and the SUDARE surveys support steeper DTDs. The origin of this discrepancy is unclear: for the red galaxies, the higher rate measured in LOSS compared to the other surveys could be ascribed to the superior statistics, but the discrepancy for the blue galaxies is puzzling. We need to understand the origin of this discrepancy to investigate the shape of the DTD, for example with a more precise characterization of the properties of the reddest and bluest galaxies concerning their intrinsic colours and SNIa rates.

To get further insight on this question we consider the correlation between the SNIa rate and the specific SFR of the parent galaxy. This allows us to include data from other independent surveys and 
to  minimize the systematics introduced by the SSP modelling.
In Fig.~\ref{fig_ria_ssfr} we show several determinations of such a correlation made on the three surveys SUDARE, CET99, and LOSS, plus measurements in the literature. For the galaxies in the SUDARE survey we use the sSFR determination output of the FAST code. For the galaxies in the LOSS and CET99 samples estimates of the SFR are not readily available, so that the corresponding points in Fig. \ref{fig_ria_ssfr} are placed exploiting the relation between colour and sSFR shown in Fig.~\ref{fig_ssfr_col}.

For the literature values as far as the choice of galaxy parameter and binning are concerned we have to use measurements as they have been published
on the original papers. \cite{Sullivan:2006nq} obtained a measurement of the rate per unit mass as a function of the galaxy sSFR using data of the Supernova Legacy Survey. Similar measurements were later provided using data from the Sloan Digital Sky Survey (SDSS) by \cite{smith:2012kx} (imaging search) and by \cite{graur:2013fd} (spectroscopic search).
The literature rates were obtained under the assumption of an IMF with a flat slope in the low mass range; we rescale these rates to our adopted Salpeter IMF by multiplying the quoted  values by a factor of 0.76.

The observations of the SNIa rate versus the specific SFR are compared in Fig.~\ref{fig_ria_ssfr} to our models for the different DTDs. There is a reasonable agreement among the various  datasets; however for the starburst galaxies the rate of the LOSS survey is relatively low whereas for the passive galaxies the CET99 rate appears particularly low. Actually, the difference between the LOSS and the CET99 rate on this plot mirrors that found from the correlation of the rate with the parent galaxy colours.  The discrepancy between the literature values
(filled symbols) and our determination for the passive galaxies in the LOSS survey may instead arise from the different evaluation of the galaxy masses used to normalize the rate. 
Indeed, as shown in Fig.~\ref{fig_data}, the way in which the galaxy masses are determined induces a systematic effect in the rate estimates, which also appears to affect the relation with the specific SFR. In Fig.~\ref{fig_ria_ssfr} we also show the results of simulation by \cite{graur+2015} based on a power law DTD. The model was tailored to the galaxies in the \cite{graur+2015} dataset, and indeed they fit the filled diamonds in Fig.~ \ref{fig_ria_ssfr}. However, the difference between the various observational measurements weakens the conclusion for this model as well.

 To summarize, once allowed for statistical and systematic uncertainties, the various measurements appear to be in broad agreement, but at present no robust conclusion on the DTD can be drawn from the comparison between the models and the data. 

\begin{figure}
\centering
\resizebox{\hsize}{!}{
\includegraphics[angle=0,clip=true]{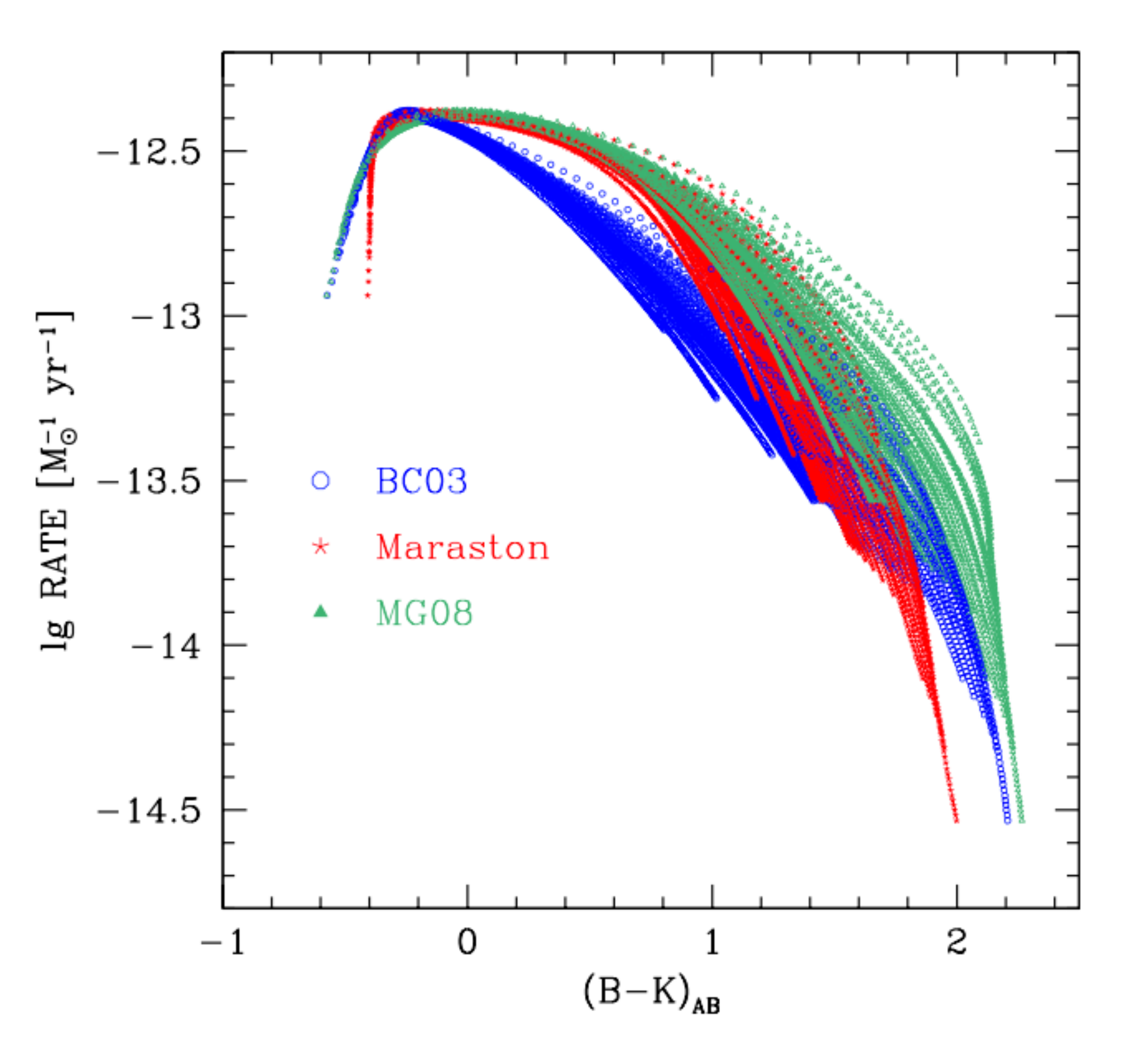}}
\caption{Theoretical correlation between the rate of SNIa and the colour of the parent galaxy for the three sets of SSP models as labelled, adopting  the  log-normal SFHs plotted in Fig. \ref{fig_abramo} and the SD DTD. }
        \label{fig_sum_ssps}
\end{figure}

\begin{figure}
\centering
\resizebox{\hsize}{!}{
\includegraphics[angle=0,clip=true]{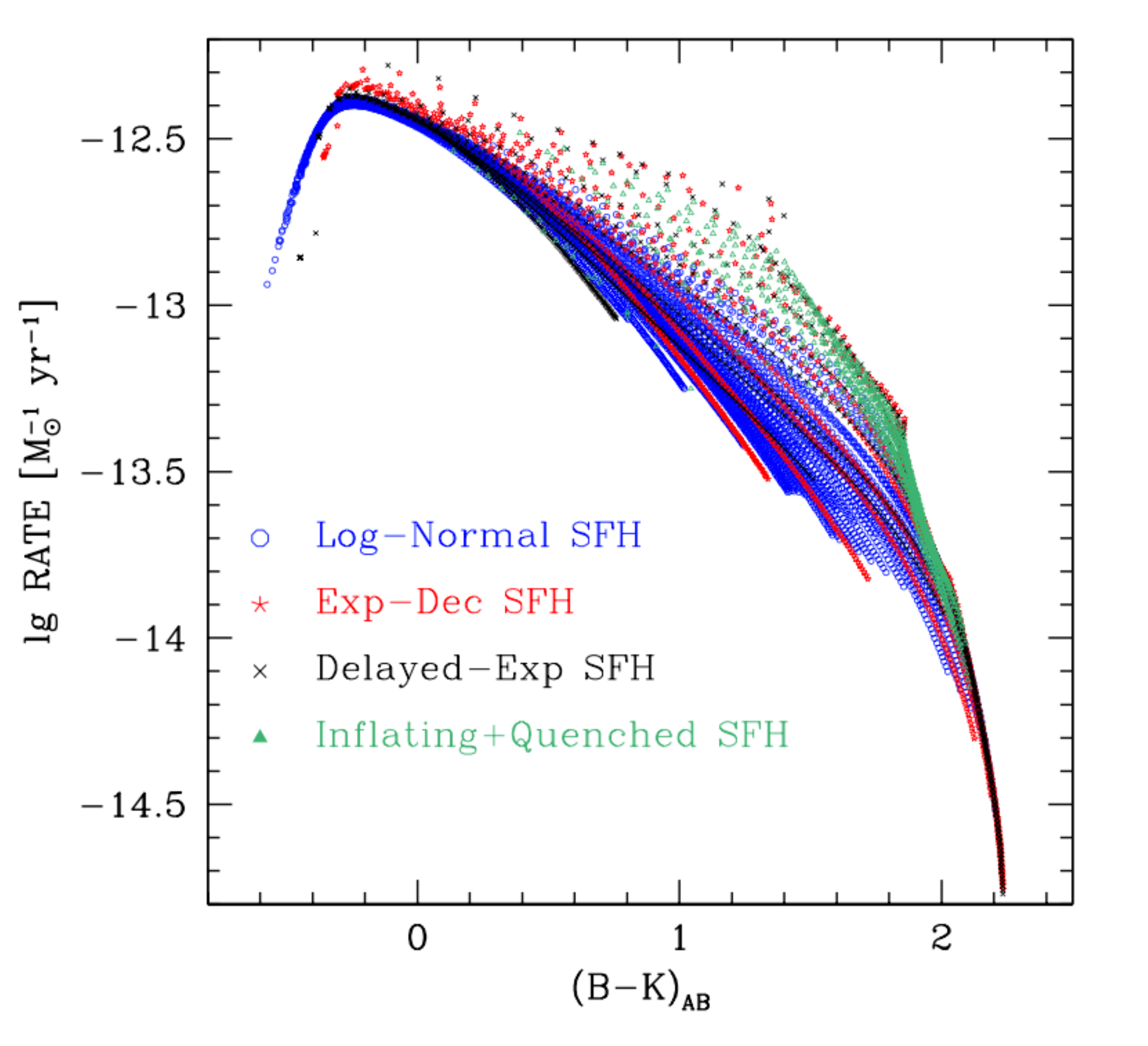}}
\caption{Theoretical correlation between the rate of SNIa and the colour of the parent galaxy for the four SFH laws and the SD DTD. Colours are based on adopting the BC03 solar metallicity SSP. }
        \label{fig_sum_sfh}
\end{figure}

\section{Summary and conclusions}

In this paper we presented  a detailed investigation of the systematic effects which hamper the derivation of the delay times distribution of SNIa progenitors from the analysis of their observed rate correlations with the properties of the parent galaxies.
Specifically, we considered the effect of different sets of SSP models to compute the  galaxy colours,  different SFH laws in the galaxy population, and  different ways to evaluate the galaxy masses when computing the rate per unit mass in observational samples. We summarize our results as follows:

\begin{itemize}
\item Different SSP sets provide different galaxy predicted colours for a given SFH. As a consequence, the correlation between the SNIa rate and the colour of the parent galaxy depends on the set of SSP models adopted, as illustrated in  Fig.~\ref{fig_sum_ssps} for one particular DTD model.  A similar result holds for the other DTDs considered in this paper. The effect also depends on the considered colour. In general, for a given observed correlation, both the derived SNIa productivity and slope of the DTD  vary with the adopted set of SSP models. It is then very important to ensure that the chosen SSP set consistently describes the properties of the galaxy sample.  

\item For a given DTD and SSP set, the models provide different renditions of the correlation between the SNIa rate and the parent galaxy colour, depending on the adopted SFH , for example the log-normal SFHs populate a narrower strip on this plane compared to the other options considered in this work (cf. Fig.~\ref{fig_sum_sfh}).  Again, a similar trend is obtained with the other DTDs considered in this paper.

\item Further uncertainty comes from the normalization of the observed rate to the galaxy mass.

When estimating the galaxy mass, systematic differences arise from (i) different choices for the IMF, (ii) whether the total mass of formed stars or the current stellar mass is considered, and (iii) the way in which the dependence of the M/L ratio on the age distribution in each galaxy is accounted for. We emphasize that to the end of deriving a robust estimate of \kia\ and of the slope of the DTD, the total mass of formed stars is the best tracer of the galaxy size, since the reduction factor due to the mass recycling in the evolution of the galaxy depends on details of the modelling (see Eq. (\ref{eq_rate_nor})). In addition, the more accurate the assessment of the age distribution in the galaxy, the  more reliable the determination of the mass-to-light ratio. Therefore, a procedure based on SED fitting on a wide colour baseline should be preferred to a plain relation between integrated colour and the mass-to-light ratio.

\item The above systematic uncertainties have to be compared with the size of the effect we want to measure.  As shown in Figures \ref{fig_corr_umj} and \ref{fig_corr_bmk}, at fixed scenario for the SFH and SSP model set, the dependence of the correlation on the DTD is not so dramatic to produce radically different trends. This follows from the similarity of the considered DTDs (see Fig. \ref{fig_dtds}) combined with the relatively wide age distribution of stars in galaxies. However, steeper DTDs predict steeper correlations, so that accurate rates measured in the bluest and in the reddest galaxies allow us to discriminate among the different models. 
\end{itemize}

The computations presented in this paper are based on DTDs characteristic of either SD or DD progenitors. However, as mentioned in the Introduction, it is more likely that both evolutionary channels are at work in nature. \cite{greggio_2010} considered two extreme possibilities for mixed models: the Solomonic mixture, in which both SD and DD channels contribute 50 \% of the explosions at any delay time; and the segregated mixture, in which SD (DD) explosions contribute all events with delay times shorter (longer) than 0.15 Gyr, and both channels provide half of the total events from one stellar generation  over a Hubble time. The theoretical correlations with the colours of the parent galaxies obtained with these two mixtures are very similar to those shown in this work, but for the segregated mixture in very blue galaxies. In fact, this mixture provides a very high fraction of prompt explosions, which cause a very high rate in stellar populations with very blue colours ($(B-K)_{AB} \sim -0.4$). At present there are no data for such blue galaxies to constrain this possibility for the DTD. In this respect, we point out that rather than validating a specific scenario for the DTD, our aim is to illustrate the limits of the discriminating power of the correlations at present. To this end we consider the representative DTD shapes derived in \cite{greggio_2005}, and neglect other propositions in the literature, for example the realizations of binary populations synthesis models.  Even in this simplified framework the constraining power of the correlations is subject to systematic effects which have not been considered so far. 

We remark that our study is limited to considering how the correlation between the SNIa rate and the parent galaxy properties vary for different SFH laws, while neglecting the impact of the distribution of the parameters characterizing the sample galaxies within each SFH scenario (e.g. AGE and $\tau$ for the standard laws). Our analysis rests on the implicit assumption that the galaxy population is characterized by a flat distribution of these parameters. As can be appreciated from Fig.~\ref{fig_sum_sfh}, the average rate in colour (or equivalently sSFR) bins is sensitive to the distribution of the SFH parameters within the galaxy population under consideration. Thus, a refined analysis requires constraining these parameters by, for example fitting the distribution of the sample galaxies on the two-colour diagram. Alternatively, fitting the SFH to each galaxy to determine the expected SNIa rate would tailor the predictions to the specific galaxy sample. Although we did not perform such fit, our analysis shows that the actual distribution of the SFH parameters introduces one further systematic in the derivation of the DTD.  

\begin{figure}
\centering
\resizebox{\hsize}{!}{
\includegraphics[angle=0,clip=true]{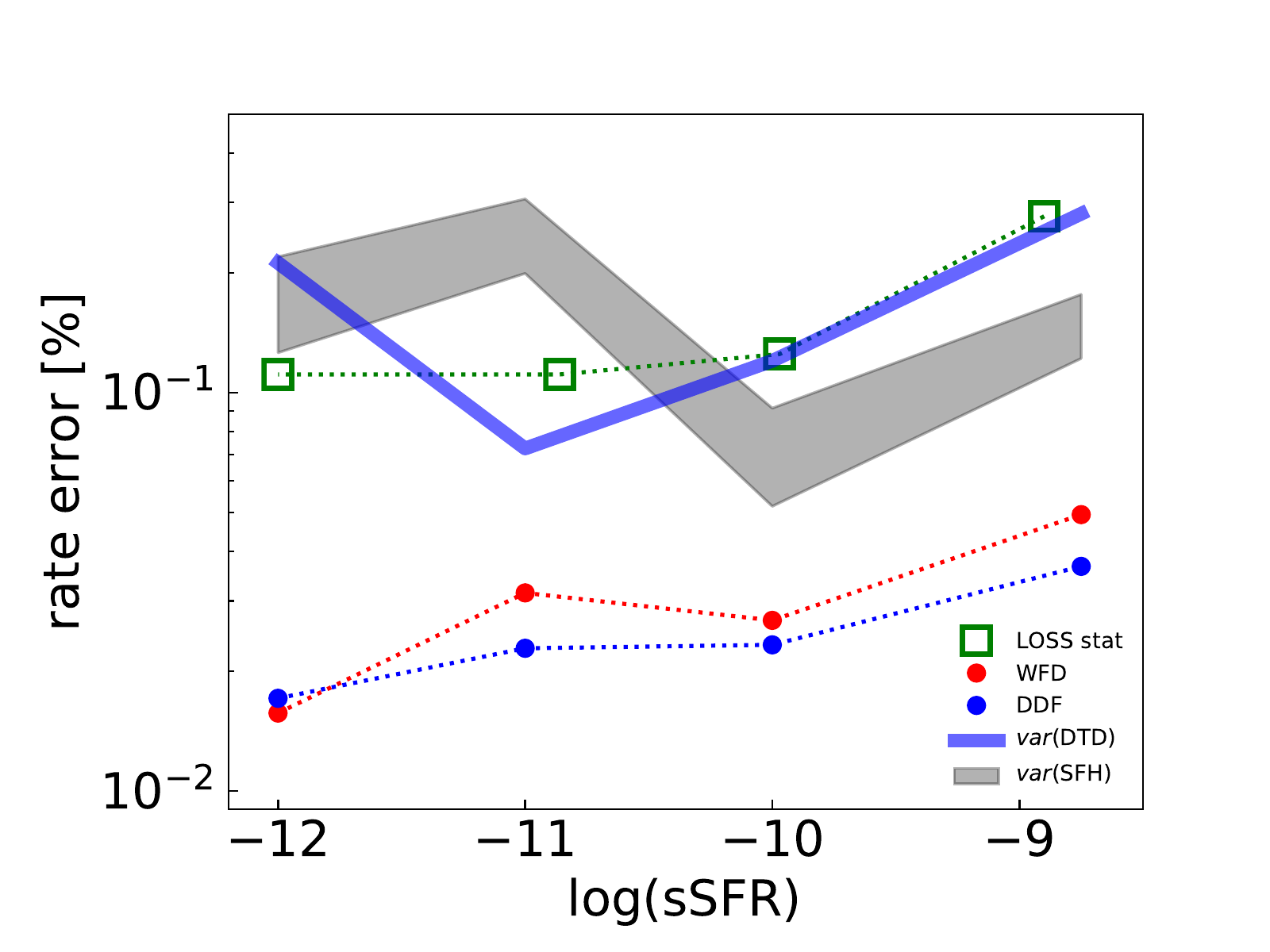}}
\caption{Theoretical and observational relative uncertainties on the value of the SNIa rate per unit mass as a function of the specific SFR. Open squares and dots show the statistical uncertainties for the LOSS survey and expected for the WFD and DDF LSST surveys, as labelled. The blue and grey stripes show the variance of the theoretical rates associated with different DTD models and different SFHs, respectively.}
        \label{fig_sum_simul}
\end{figure}

The discussion above leads to the key question of, given the many statistical and systematic uncertainties, what the prospects are for using the existing rate measurements and those expected from the next wide field surveys to discriminate among alternative DTD scenarios.

To evaluate the constraining power of current dataset, we used the information from three SN surveys (SUDARE, LOSS, and CET99). As reference model, we chose the log-normal SFH description, the BC03 SSP models at solar metallicity,  and a single slope Salpeter IMF from 0.1 to 100 \msun. With these parameters we find that adopting an average value $\kia \simeq 0.8 \times 10^{-3} \msun^{-1}$ broadly accounts for the observed correlations. This value is in excellent agreement with that found in Paper I from the analysis of the cosmic SNIa rate. Yet, the current data do not allow us to discriminate among the different DTD because of the discrepancies among the three datasets. The LOSS data indicate a flat DTD that is compatible with the DDW model; the CET99 and the SUDARE data indicate steeper DTDs. We conclude that at present all the DTDs considered in this work, which span a great variety of models for the SNIa progenitors, are consistent with the observations once allowing for both the statistical and systematic errors. 

 In an attempt to quantify the effect that we want to measure and compare this effect with the different contributions to the uncertainties, in Fig.~\ref{fig_sum_simul} we plot the variance (standard deviation) of the rates predicted by the different DTD models (SD, DDC, DDW, and P.L.) in bins of specific SFR (blue strip). This is intended to represent the average difference between DTD models. Clearly, if we want to be able to discriminate between the various DTDs, the observational errors need to be smaller that this level.
 As we mentioned above, at present this is not the case. Even for the LOSS survey, which has the largest number of events, the statistical errors alone have the same magnitude as the predicted model differences (see empty squares in Fig. \ref{fig_sum_simul}). 
 
 However, there are a few ongoing or planned searches that have the potential to strongly increase the number of detected events and hence lower the statistical error. Among these,
 the upcoming LSST survey will provide the largest database for the SNIa ever, i.e. possibly up to a million events in the whole  ten years of operation.  Yet, not all the transients will have sufficient data to be properly classified.  We use the conservative estimates derived in \cite{astier_2014} for  the number of SNIa discovered and reliably classified in the LSST Wide Field Survey (WFS) and in the Deep Drilling Fields (DDF), which count about $10^4$ events each. Fig.~\ref{fig_sum_simul} shows the statistical uncertainties in the different sSFR bins associated with the \cite{astier_2014} estimates (red and blue dots). It can be seen that the predicted number of events will be so high in all sSFR bins to bring the statistical uncertainty much below the requirements to discriminate different DTDs. It is worth noting that the recent simulations of the LSST DESC collaboration carried out with the baseline survey strategy show that the survey will provide a superb statistics, with $\gtrsim$ 100000 SNIa with well sampled light curves and spectroscopic redshift
 \citep{DESC_2018}.
 At this point we need to turn our attention to the systematic errors.

According to \cite{Li_2011} the systematic error affecting their observed rates is of a similar magnitude as the statistical error, but it is mostly related to the choice of the global parameters of the computation (cf. Sect. 3.4). Therefore, while this error component is relevant for the determination of \kia, it should not impact on the trend of the rate as a function of the galaxy parameters, for which only the statistical errors should be considered. 

On the other hand, the grey strip in Fig.~\ref{fig_sum_simul} shows the variance of the rates for the four different SFH prescriptions explored in this work.
It appears that differencies in the SFH and DTD induce a comparable variance of the rate, so that ambiguities on the SFH  can potentially blur the chance to identify the dominant SNIa model.
 
 Therefore, in spite of the  excellent statistics which will be available with LSST data, the conclusions on the DTD will be affected by systematic uncertainties related to the SFH in the galaxies, and robust results will be achievable only with well constrained  age distributions in the sample objects.
 The LSST dataset will also contain homogeneous multicolour data for the survey galaxies, which will allow a fair characterization of the stellar populations, providing an ideal dataset for this kind of investigation. 
 
To conclude, in order to reach firm constraints on the slope of the DTD from the correlation of the SNIa rate with the properties of the parent galaxy, we need

\begin{itemize}
\item large galaxy samples, especially at the bluest and reddest colours; this will decrease the statistical uncertainty at the ends of the correlation, improving the leverage over the slope of the DTD;
\item accurate measurements of redshift and extinction to be able to analyse the correlations in the rest-frame dereddened colours;
\item well-constrained age distributions and masses of the sample galaxies, from multiband photometry ranging from the UV to the IR, i. e. sensitive to the whole range of stellar ages.
\end{itemize}

The LSST survey appears very promising  in this respect, providing an unbiased, homogeneous, and vast database. In parallel, we need to develop theoretical and observational studies on  the SFH in galaxies, which account for the various properties of big samples both in the local and  distant Universe. This will enable us to describe at best the distribution of stellar ages of the sample objects, a necessary ingredient to the end of constraining the DTD from the analysis of the correlation of the SNIa rate with the parent galaxy properties.

\begin{acknowledgements}
We warmly thank Maria Teresa Botticella for fruitful discussions on the scientific content of the paper, and for a careful reading of the manuscript.
\end{acknowledgements}

\bibliographystyle{aa}
\bibliography{paper.bib}

\begin{thebibliography}{71}
\expandafter\ifx\csname natexlab\endcsname\relax\def\natexlab#1{#1}\fi

\bibitem[{{Abramson} {et~al.}(2016){Abramson}, {Gladders}, {Dressler},
  {Oemler}, {Poggianti}, \& {Vulcani}}]{abramo_2016}
{Abramson}, L.~E., {Gladders}, M.~D., {Dressler}, A., {et~al.} 2016, \apj, 832,
  7

\bibitem[{{Astier} {et~al.}(2014){Astier}, {Balland}, {Brescia}, {Cappellaro},
  {Carlberg}, {Cavuoti}, {Della Valle}, {Gangler}, {Goobar}, {Guy}, {Hardin},
  {Hook}, {Kessler}, {Kim}, {Linder}, {Longo}, {Maguire}, {Mannucci},
  {Mattila}, {Nichol}, {Pain}, {Regnault}, {Spiro}, {Sullivan}, {Tao},
  {Turatto}, {Wang}, \& {Wood-Vasey}}]{astier_2014}
{Astier}, P., {Balland}, C., {Brescia}, M., {et~al.} 2014, \aap, 572, A80

\bibitem[{{Blanc} \& {Greggio}(2008)}]{blanc:2008rp}
{Blanc}, G. \& {Greggio}, L. 2008, \na, 13, 606

\bibitem[{{Botticella} {et~al.}(2017){Botticella}, {Cappellaro}, {Greggio},
  {Pignata}, {Della Valle}, {Grado}, {Limatola}, {Baruffolo}, {Benetti},
  {Bufano}, {Capaccioli}, {Cascone}, {Covone}, {De Cicco}, {Falocco},
  {Haeussler}, {Harutyunyan}, {Jarvis}, {Marchetti}, {Napolitano}, {Paolillo},
  {Pastorello}, {Radovich}, {Schipani}, {Tomasella}, {Turatto}, \&
  {Vaccari}}]{Botticella_2017}
{Botticella}, M.~T., {Cappellaro}, E., {Greggio}, L., {et~al.} 2017, \aap, 598,
  A50

\bibitem[{{Bruzual} \& {Charlot}(2003)}]{Bruzual:2003kx}
{Bruzual}, G. \& {Charlot}, S. 2003, \mnras, 344, 1000

\bibitem[{{Calzetti}(2001)}]{calzetti_2001}
{Calzetti}, D. 2001, \pasp, 113, 1449

\bibitem[{{Cao} {et~al.}(2015){Cao}, {Kulkarni}, {Howell}, {Gal-Yam},
  {Kasliwal}, {Valenti}, {Johansson}, {Amanullah}, {Goobar}, {Sollerman},
  {Taddia}, {Horesh}, {Sagiv}, {Cenko}, {Nugent}, {Arcavi}, {Surace},
  {Wo{\'z}niak}, {Moody}, {Rebbapragada}, {Bue}, \& {Gehrels}}]{cao_2015}
{Cao}, Y., {Kulkarni}, S.~R., {Howell}, D.~A., {et~al.} 2015, \nat, 521, 328

\bibitem[{{Cappellaro} {et~al.}(2015){Cappellaro}, {Botticella}, {Pignata},
  {Grado}, {Greggio}, {Limatola}, {Vaccari}, {Baruffolo}, {Benetti}, {Bufano},
  {Capaccioli}, {Cascone}, {Covone}, {De Cicco}, {Falocco}, {Della Valle},
  {Jarvis}, {Marchetti}, {Napolitano}, {Paolillo}, {Pastorello}, {Radovich},
  {Schipani}, {Spiro}, {Tomasella}, \& {Turatto}}]{cappellaro+2015}
{Cappellaro}, E., {Botticella}, M.~T., {Pignata}, G., {et~al.} 2015, \aap, 584,
  A62

\bibitem[{{Cappellaro} {et~al.}(1999){Cappellaro}, {Evans}, \&
  {Turatto}}]{cappellaro:1999dg}
{Cappellaro}, E., {Evans}, R., \& {Turatto}, M. 1999, \aap, 351, 459

\bibitem[{{Cardelli} {et~al.}(1989){Cardelli}, {Clayton}, \&
  {Mathis}}]{cardelli_1989}
{Cardelli}, J.~A., {Clayton}, G.~C., \& {Mathis}, J.~S. 1989, \apj, 345, 245

\bibitem[{{Childress} {et~al.}(2013){Childress}, {Aldering}, {Antilogus},
  {Aragon}, {Bailey}, {Baltay}, {Bongard}, {Buton}, {Canto}, {Cellier-Holzem},
  {Chotard}, {Copin}, {Fakhouri}, {Gangler}, {Guy}, {Hsiao}, {Kerschhaggl},
  {Kim}, {Kowalski}, {Loken}, {Nugent}, {Paech}, {Pain}, {Pecontal}, {Pereira},
  {Perlmutter}, {Rabinowitz}, {Rigault}, {Runge}, {Scalzo}, {Smadja}, {Tao},
  {Thomas}, {Weaver}, \& {Wu}}]{Childress_2013}
{Childress}, M., {Aldering}, G., {Antilogus}, P., {et~al.} 2013, \apj, 770, 108

\bibitem[{{Chiosi} {et~al.}(2017){Chiosi}, {Sciarratta}, {D'Onofrio}, {Chiosi},
  {Brotto}, {De Michele}, \& {Politino}}]{chiosi_2017}
{Chiosi}, C., {Sciarratta}, M., {D'Onofrio}, M., {et~al.} 2017, \apj, 851, 44

\bibitem[{{Ciotti} {et~al.}(1991){Ciotti}, {D'Ercole}, {Pellegrini}, \&
  {Renzini}}]{Ciotti_1991}
{Ciotti}, L., {D'Ercole}, A., {Pellegrini}, S., \& {Renzini}, A. 1991, \apj,
  376, 380

\bibitem[{{Claeys} {et~al.}(2014){Claeys}, {Pols}, {Izzard}, {Vink}, \&
  {Verbunt}}]{Claeys_2014}
{Claeys}, J.~S.~W., {Pols}, O.~R., {Izzard}, R.~G., {Vink}, J., \& {Verbunt},
  F.~W.~M. 2014, \aap, 563, A83

\bibitem[{{Dimitriadis} {et~al.}(2019){Dimitriadis}, {Foley}, {Rest}, {Kasen},
  {Piro}, {Polin}, {Jones}, {Villar}, {Narayan}, {Coulter}, {Kilpatrick},
  {Pan}, {Rojas-Bravo}, {Fox}, {Jha}, {Nugent}, {Riess}, {Scolnic}, {Drout},
  {K2 Mission Team}, {Barentsen}, {Dotson}, {Gully-Santiago}, {Hedges}, {Cody},
  {Barclay}, {Howell}, {KEGS}, {Garnavich}, {Tucker}, {Shaya}, {Mushotzky},
  {Olling}, {Margheim}, {Zenteno}, {Kepler spacecraft team}, {Coughlin}, {Van
  Cleve}, {Cardoso}, {Larson}, {McCalmont-Everton}, {Peterson}, {Ross},
  {Reedy}, {Osborne}, {McGinn}, {Kohnert}, {Migliorini}, {Wheaton}, {Spencer},
  {Labonde}, {Castillo}, {Beerman}, {Steward}, {Hanley}, {Larsen},
  {Gangopadhyay}, {Kloetzel}, {Weschler}, {Nystrom}, {Moffatt}, {Redick},
  {Griest}, {Packard}, {Muszynski}, {Kampmeier}, {Bjella}, {Flynn},
  {Elsaesser}, {Pan-STARRS}, {Chambers}, {Flewelling}, {Huber}, {Magnier},
  {Waters}, {Schultz}, {Bulger}, {Lowe}, {Willman}, {Smartt}, {Smith}, {DECam},
  {Points}, {Strampelli}, {ASAS-SN}, {Brimacombe}, {Chen}, {Mu{\~n}oz},
  {Mutel}, {Shields}, {Vallely}, {Villanueva}, {PTSS/TNTS}, {Li}, {Wang},
  {Zhang}, {Lin}, {Mo}, {Zhao}, {Sai}, {Zhang}, {Zhang}, {Zhang}, {Wang},
  {Zhang}, {Baron}, {DerKacy}, {Li}, {Chen}, {Xiang}, {Rui}, {Wang}, {Huang},
  {Li}, {Cumbres Observatory}, {Hosseinzadeh}, {Howell}, {Arcavi}, {Hiramatsu},
  {Burke}, {Valenti}, {ATLAS}, {Tonry}, {Denneau}, {Heinze}, {Weiland},
  {Stalder}, {Konkoly}, {Vink{\'o}}, {S{\'a}rneczky}, {P{\'a}l}, {B{\'o}di},
  {Bogn{\'a}r}, {Cs{\'a}k}, {Cseh}, {Cs{\"o}rnyei}, {Hanyecz}, {Ign{\'a}cz},
  {Kalup}, {K{\"o}nyves-T{\'o}th}, {Kriskovics}, {Ordasi}, {Rajmon},
  {S{\'o}dor}, {Szab{\'o}}, {Szak{\'a}ts}, {Zsidi}, {ePESSTO}, {Williams},
  {Nordin}, {Cartier}, {Frohmaier}, {Galbany}, {Guti{\'e}rrez}, {Hook},
  {Inserra}, {Smith}, {Arizona}, {Sand}, {Andrews}, {Smith}, \&
  {Bilinski}}]{Dimitriadis_2019}
{Dimitriadis}, G., {Foley}, R.~J., {Rest}, A., {et~al.} 2019, \apjl, 870, L1

\bibitem[{{Eales} {et~al.}(2018){Eales}, {Smith}, {Bourne}, {Loveday},
  {Rowlands}, {van der Werf}, {Driver}, {Dunne}, {Dye}, {Furlanetto}, {Ivison},
  {Maddox}, {Robotham}, {Smith}, {Taylor}, {Valiante}, {Wright}, {Cigan}, {De
  Zotti}, {Jarvis}, {Marchetti}, {Micha{\l}owski}, {Phillipps}, {Viaene}, \&
  {Vlahakis}}]{Eales_2018}
{Eales}, S., {Smith}, D., {Bourne}, N., {et~al.} 2018, \mnras, 473, 3507

\bibitem[{{Eldridge} {et~al.}(2017){Eldridge}, {Stanway}, {Xiao}, {McClelland
  }, {Taylor}, {Ng}, {Greis}, \& {Bray}}]{Eldridge_2017}
{Eldridge}, J.~J., {Stanway}, E.~R., {Xiao}, L., {et~al.} 2017, Publications of
  the Astronomical Society of Australia, 34, e058

\bibitem[{{Farmer} \& {Phinney}(2003)}]{Farmer_2003}
{Farmer}, A.~J. \& {Phinney}, E.~S. 2003, \mnras, 346, 1197

\bibitem[{{Gallazzi} {et~al.}(2005){Gallazzi}, {Charlot}, {Brinchmann},
  {White}, \& {Tremonti}}]{gallazzi_2005}
{Gallazzi}, A., {Charlot}, S., {Brinchmann}, J., {White}, S.~D.~M., \&
  {Tremonti}, C.~A. 2005, \mnras, 362, 41

\bibitem[{{Gavazzi} {et~al.}(2002){Gavazzi}, {Bonfanti}, {Sanvito}, {Boselli},
  \& {Scodeggio}}]{gavazzi_2002}
{Gavazzi}, G., {Bonfanti}, C., {Sanvito}, G., {Boselli}, A., \& {Scodeggio}, M.
  2002, \apj, 576, 135

\bibitem[{{Gladders} {et~al.}(2013){Gladders}, {Oemler}, {Dressler},
  {Poggianti}, {Vulcani}, \& {Abramson}}]{Gladders_2013}
{Gladders}, M.~D., {Oemler}, A., {Dressler}, A., {et~al.} 2013, \apj, 770, 64

\bibitem[{{Gonz{\'a}lez Hern{\'a}ndez} {et~al.}(2012){Gonz{\'a}lez
  Hern{\'a}ndez}, {Ruiz-Lapuente}, {Tabernero}, {Montes}, {Canal},
  {M{\'e}ndez}, \& {Bedin}}]{gonzales_2012}
{Gonz{\'a}lez Hern{\'a}ndez}, J.~I., {Ruiz-Lapuente}, P., {Tabernero}, H.~M.,
  {et~al.} 2012, \nat, 489, 533

\bibitem[{{Graur} {et~al.}(2017){Graur}, {Bianco}, {Huang}, {Modjaz},
  {Shivvers}, {Filippenko}, {Li}, \& {Eldridge}}]{Graur_2017}
{Graur}, O., {Bianco}, F.~B., {Huang}, S., {et~al.} 2017, \apj, 837, 120

\bibitem[{{Graur} {et~al.}(2015){Graur}, {Bianco}, \& {Modjaz}}]{graur+2015}
{Graur}, O., {Bianco}, F.~B., \& {Modjaz}, M. 2015, \mnras, 450, 905

\bibitem[{{Graur} \& {Maoz}(2013)}]{graur:2013fd}
{Graur}, O. \& {Maoz}, D. 2013, \mnras, 430, 1746

\bibitem[{{Greggio}(2005)}]{greggio_2005}
{Greggio}, L. 2005, \aap, 441, 1055

\bibitem[{{Greggio}(2010)}]{greggio_2010}
{Greggio}, L. 2010, \mnras, 406, 22

\bibitem[{{Greggio} \& {Cappellaro}(2009)}]{greggio+2009}
{Greggio}, L. \& {Cappellaro}, E. 2009, in American Institute of Physics
  Conference Series, Vol. 1111, American Institute of Physics Conference
  Series, ed. G.~{Giobbi}, A.~{Tornambe}, G.~{Raimondo}, M.~{Limongi}, L.~A.
  {Antonelli}, N.~{Menci}, \& E.~{Brocato}, 477--484

\bibitem[{{Greggio} \& {Renzini}(2011)}]{greggio_book}
{Greggio}, L. \& {Renzini}, A. 2011, {Stellar Populations. A User Guide from
  Low to High Redshift} (Wiley-VCH-Verlag)

\bibitem[{{Greggio} {et~al.}(2008){Greggio}, {Renzini}, \& {Daddi}}]{GRD08}
{Greggio}, L., {Renzini}, A., \& {Daddi}, E. 2008, \mnras, 388, 829

\bibitem[{{Hillebrandt} {et~al.}(2013){Hillebrandt}, {Kromer}, {R{\"o}pke}, \&
  {Ruiter}}]{Hillebrandt_2013}
{Hillebrandt}, W., {Kromer}, M., {R{\"o}pke}, F.~K., \& {Ruiter}, A.~J. 2013,
  Frontiers of Physics, 8, 116

\bibitem[{{Hosseinzadeh} {et~al.}(2017){Hosseinzadeh}, {Sand}, {Valenti},
  {Brown}, {Howell}, {McCully}, {Kasen}, {Arcavi}, {Azalee Bostroem},
  {Tartaglia}, {Hsiao}, {Davis}, {Shahbandeh}, \&
  {Stritzinger}}]{Hosseinzadeh_2017}
{Hosseinzadeh}, G., {Sand}, D.~J., {Valenti}, S., {et~al.} 2017, \apj, 845, L11

\bibitem[{{Iben} \& {Tutukov}(1984)}]{IT_1984}
{Iben}, Jr., I. \& {Tutukov}, A.~V. 1984, \apjs, 54, 335

\bibitem[{{Kobayashi} {et~al.}(2015){Kobayashi}, {Nomoto}, \&
  {Hachisu}}]{Kobayashi_2015}
{Kobayashi}, C., {Nomoto}, K., \& {Hachisu}, I. 2015, \apjl, 804, L24

\bibitem[{{Kouwenhoven} {et~al.}(2007){Kouwenhoven}, {Brown}, {Portegies
  Zwart}, \& {Kaper}}]{Kouwenhoven_2007}
{Kouwenhoven}, M.~B.~N., {Brown}, A.~G.~A., {Portegies Zwart}, S.~F., \&
  {Kaper}, L. 2007, \aap, 474, 77

\bibitem[{{Kriek} {et~al.}(2009){Kriek}, {van Dokkum}, {Labb{\'e}}, {Franx},
  {Illingworth}, {Marchesini}, \& {Quadri}}]{Kriek2009}
{Kriek}, M., {van Dokkum}, P.~G., {Labb{\'e}}, I., {et~al.} 2009, \apj, 700,
  221

\bibitem[{{Li} {et~al.}(2011{\natexlab{a}}){Li}, {Bloom}, {Podsiadlowski},
  {Miller}, {Cenko}, {Jha}, {Sullivan}, {Howell}, {Nugent}, {Butler}, {Ofek},
  {Kasliwal}, {Richards}, {Stockton}, {Shih}, {Bildsten}, {Shara}, {Bibby},
  {Filippenko}, {Ganeshalingam}, {Silverman}, {Kulkarni}, {Law}, {Poznanski},
  {Quimby}, {McCully}, {Patel}, {Maguire}, \& {Shen}}]{Li_2011Nature}
{Li}, W., {Bloom}, J.~S., {Podsiadlowski}, P., {et~al.} 2011{\natexlab{a}},
  \nat, 480, 348

\bibitem[{{Li} {et~al.}(2011{\natexlab{b}}){Li}, {Chornock}, {Leaman},
  {Filippenko}, {Poznanski}, {Wang}, {Ganeshalingam}, \& {Mannucci}}]{Li_2011}
{Li}, W., {Chornock}, R., {Leaman}, J., {et~al.} 2011{\natexlab{b}}, \mnras,
  412, 1473

\bibitem[{{Madau} \& {Dickinson}(2014)}]{madau:2014uf}
{Madau}, P. \& {Dickinson}, M. 2014, \araa, 52, 415

\bibitem[{{Maeda} \& {Terada}(2016)}]{Maeda_2016}
{Maeda}, K. \& {Terada}, Y. 2016, International Journal of Modern Physics D,
  25, 1630024

\bibitem[{{Mannucci} {et~al.}(2006){Mannucci}, {Della Valle}, \&
  {Panagia}}]{mannucci:2006zi}
{Mannucci}, F., {Della Valle}, M., \& {Panagia}, N. 2006, \mnras, 370, 773

\bibitem[{{Mannucci} {et~al.}(2005){Mannucci}, {Della Valle}, {Panagia},
  {Cappellaro}, {Cresci}, {Maiolino}, {Petrosian}, \&
  {Turatto}}]{Mannucci_2005}
{Mannucci}, F., {Della Valle}, M., {Panagia}, N., {et~al.} 2005, \aap, 433, 807

\bibitem[{{Maoz} \& {Gal-Yam}(2004)}]{Maoz_2004}
{Maoz}, D. \& {Gal-Yam}, A. 2004, \mnras, 347, 951

\bibitem[{{Maoz} \& {Mannucci}(2012)}]{MM_2012}
{Maoz}, D. \& {Mannucci}, F. 2012, \pasa, 29, 447

\bibitem[{{Maoz} {et~al.}(2012){Maoz}, {Mannucci}, \& {Brandt}}]{Maoz_2012}
{Maoz}, D., {Mannucci}, F., \& {Brandt}, T.~D. 2012, \mnras, 426, 3282

\bibitem[{{Maoz} {et~al.}(2011){Maoz}, {Mannucci}, {Li}, {Filippenko}, {Della
  Valle}, \& {Panagia}}]{maoz+2011}
{Maoz}, D., {Mannucci}, F., {Li}, W., {et~al.} 2011, \mnras, 412, 1508

\bibitem[{{Maraston} {et~al.}(2010){Maraston}, {Pforr}, {Renzini}, {Daddi},
  {Dickinson}, {Cimatti}, \& {Tonini}}]{maraston+2010}
{Maraston}, C., {Pforr}, J., {Renzini}, A., {et~al.} 2010, \mnras, 407, 830

\bibitem[{{Marigo} {et~al.}(2008){Marigo}, {Girardi}, {Bressan}, {Groenewegen},
  {Silva}, \& {Granato}}]{Marigo_2008}
{Marigo}, P., {Girardi}, L., {Bressan}, A., {et~al.} 2008, \aap, 482, 883

\bibitem[{{Matteucci} \& {Greggio}(1986)}]{matteucci:1986kq}
{Matteucci}, F. \& {Greggio}, L. 1986, \aap, 154, 279

\bibitem[{{Mennekens} {et~al.}(2010){Mennekens}, {Vanbeveren}, {De Greve}, \&
  {De Donder}}]{mennekens_2010}
{Mennekens}, N., {Vanbeveren}, D., {De Greve}, J.~P., \& {De Donder}, E. 2010,
  \aap, 515, A89

\bibitem[{{Oemler} {et~al.}(2013){Oemler}, {Dressler}, {Gladders}, {Fritz},
  {Poggianti}, {Vulcani}, \& {Abramson}}]{Oemler_2013}
{Oemler}, Jr., A., {Dressler}, A., {Gladders}, M.~G., {et~al.} 2013, \apj, 770,
  63

\bibitem[{{Patat} {et~al.}(2007){Patat}, {Chandra}, {Chevalier}, {Justham},
  {Podsiadlowski}, {Wolf}, {Gal-Yam}, {Pasquini}, {Crawford}, {Mazzali},
  {Pauldrach}, {Nomoto}, {Benetti}, {Cappellaro}, {Elias-Rosa}, {Hillebrandt},
  {Leonard}, {Pastorello}, {Renzini}, {Sabbadin}, {Simon}, \&
  {Turatto}}]{patat:2007sh}
{Patat}, F., {Chandra}, P., {Chevalier}, R., {et~al.} 2007, Science, 317, 924

\bibitem[{{Peng} {et~al.}(2010){Peng}, {Lilly}, {Kova{\v c}}, {Bolzonella},
  {Pozzetti}, {Renzini}, {Zamorani}, {Ilbert}, {Knobel}, {Iovino}, {Maier},
  {Cucciati}, {Tasca}, {Carollo}, {Silverman}, {Kampczyk}, {de Ravel},
  {Sanders}, {Scoville}, {Contini}, {Mainieri}, {Scodeggio}, {Kneib}, {Le
  F{\`e}vre}, {Bardelli}, {Bongiorno}, {Caputi}, {Coppa}, {de la Torre},
  {Franzetti}, {Garilli}, {Lamareille}, {Le Borgne}, {Le Brun}, {Mignoli},
  {Perez Montero}, {Pello}, {Ricciardelli}, {Tanaka}, {Tresse}, {Vergani},
  {Welikala}, {Zucca}, {Oesch}, {Abbas}, {Barnes}, {Bordoloi}, {Bottini},
  {Cappi}, {Cassata}, {Cimatti}, {Fumana}, {Hasinger}, {Koekemoer},
  {Leauthaud}, {Maccagni}, {Marinoni}, {McCracken}, {Memeo}, {Meneux}, {Nair},
  {Porciani}, {Presotto}, \& {Scaramella}}]{Peng_2010}
{Peng}, Y.-j., {Lilly}, S.~J., {Kova{\v c}}, K., {et~al.} 2010, \apj, 721, 193

\bibitem[{{Perlmutter} {et~al.}(1999){Perlmutter}, {Turner}, \&
  {White}}]{Perl_1999}
{Perlmutter}, S., {Turner}, M.~S., \& {White}, M. 1999, Physical Review
  Letters, 83, 670

\bibitem[{{Pritchet} {et~al.}(2008){Pritchet}, {Howell}, \&
  {Sullivan}}]{pritchet_2008}
{Pritchet}, C.~J., {Howell}, D.~A., \& {Sullivan}, M. 2008, \apjl, 683, L25

\bibitem[{{Renzini}(2009)}]{Renzini_2009}
{Renzini}, A. 2009, \mnras, 398, L58

\bibitem[{{Renzini}(2016)}]{renzini_2016}
{Renzini}, A. 2016, \mnras, 460, L45

\bibitem[{{Riess} {et~al.}(1998){Riess}, {Filippenko}, {Challis},
  {Clocchiatti}, {Diercks}, {Garnavich}, {Gilliland}, {Hogan}, {Jha},
  {Kirshner}, {Leibundgut}, {Phillips}, {Reiss}, {Schmidt}, {Schommer},
  {Smith}, {Spyromilio}, {Stubbs}, {Suntzeff}, \& {Tonry}}]{riess:1998qm}
{Riess}, A.~G., {Filippenko}, A.~V., {Challis}, P., {et~al.} 1998, \aj, 116,
  1009

\bibitem[{{Ruiter} {et~al.}(2009){Ruiter}, {Belczynski}, \&
  {Fryer}}]{Ruiter_2009}
{Ruiter}, A.~J., {Belczynski}, K., \& {Fryer}, C. 2009, \apj, 699, 2026

\bibitem[{{Ruiter} {et~al.}(2013){Ruiter}, {Sim}, {Pakmor}, {Kromer},
  {Seitenzahl}, {Belczynski}, {Fink}, {Herzog}, {Hillebrandt}, {R{\"o}pke}, \&
  {Taubenberger}}]{Ruiter_2013}
{Ruiter}, A.~J., {Sim}, S.~A., {Pakmor}, R., {et~al.} 2013, \mnras, 429, 1425

\bibitem[{{Scannapieco} \& {Bildsten}(2005)}]{scannapieco:2005rr}
{Scannapieco}, E. \& {Bildsten}, L. 2005, \apjl, 629, L85

\bibitem[{{Smartt}(2009)}]{smartt:2009mq}
{Smartt}, S.~J. 2009, \araa, 47, 63

\bibitem[{{Smith} {et~al.}(2012){Smith}, {Nichol}, {Dilday}, {Marriner},
  {Kessler}, {Bassett}, {Cinabro}, {Frieman}, {Garnavich}, {Jha}, {Lampeitl},
  {Sako}, {Schneider}, \& {Sollerman}}]{smith:2012kx}
{Smith}, M., {Nichol}, R.~C., {Dilday}, B., {et~al.} 2012, \apj, 755, 61

\bibitem[{{Stanway} \& {Eldridge}(2018)}]{Stanway_2018}
{Stanway}, E.~R. \& {Eldridge}, J.~J. 2018, \mnras, 479, 75

\bibitem[{{Sullivan} {et~al.}(2006){Sullivan}, {Le Borgne}, {Pritchet},
  {Hodsman}, {Neill}, {Howell}, {Carlberg}, {Astier}, {Aubourg}, {Balam},
  {Basa}, {Conley}, {Fabbro}, {Fouchez}, {Guy}, {Hook}, {Pain},
  {Palanque-Delabrouille}, {Perrett}, {Regnault}, {Rich}, {Taillet}, {Baumont},
  {Bronder}, {Ellis}, {Filiol}, {Lusset}, {Perlmutter}, {Ripoche}, \&
  {Tao}}]{Sullivan:2006nq}
{Sullivan}, M., {Le Borgne}, D., {Pritchet}, C.~J., {et~al.} 2006, \apj, 648,
  868

\bibitem[{{The LSST Dark Energy Science Collaboration} {et~al.}(2018){The LSST
  Dark Energy Science Collaboration}, {Mandelbaum}, {Eifler}, {Hlo{\v z}ek},
  {Collett}, {Gawiser}, {Scolnic}, {Alonso}, {Awan}, {Biswas}, {Blazek},
  {Burchat}, {Chisari}, {Dell'Antonio}, {Digel}, {Frieman}, {Goldstein},
  {Hook}, {Ivezi{\'c}}, {Kahn}, {Kamath}, {Kirkby}, {Kitching}, {Krause},
  {Leget}, {Marshall}, {Meyers}, {Miyatake}, {Newman}, {Nichol}, {Rykoff},
  {Sanchez}, {Slosar}, {Sullivan}, \& {Troxel}}]{DESC_2018}
{The LSST Dark Energy Science Collaboration}, {Mandelbaum}, R., {Eifler}, T.,
  {et~al.} 2018, arXiv e-prints [\eprint[arXiv]{1809.01669}]

\bibitem[{{Thomas} {et~al.}(1999){Thomas}, {Greggio}, \& {Bender}}]{TGB_1999}
{Thomas}, D., {Greggio}, L., \& {Bender}, R. 1999, \mnras, 302, 537

\bibitem[{{Totani} {et~al.}(2008){Totani}, {Morokuma}, {Oda}, {Doi}, \&
  {Yasuda}}]{Totani_2008}
{Totani}, T., {Morokuma}, T., {Oda}, T., {Doi}, M., \& {Yasuda}, N. 2008,
  \pasj, 60, 1327

\bibitem[{{Wang} \& {Han}(2012)}]{wang_2012}
{Wang}, B. \& {Han}, Z. 2012, New Astronomy Reviews, 56, 122

\bibitem[{{Webbink}(1984)}]{Webbink_1984}
{Webbink}, R.~F. 1984, \apj, 277, 355

\bibitem[{{Whelan} \& {Iben}(1973)}]{whelan:1973nr}
{Whelan}, J. \& {Iben}, Jr., I. 1973, \apj, 186, 1007

\end{thebibliography}

\begin{appendix} 

\section{Dependence of the model galaxy colours on the SSP set}

\begin{figure*}
\centering
\resizebox{\hsize}{!}{
\includegraphics[angle=0,clip=true]{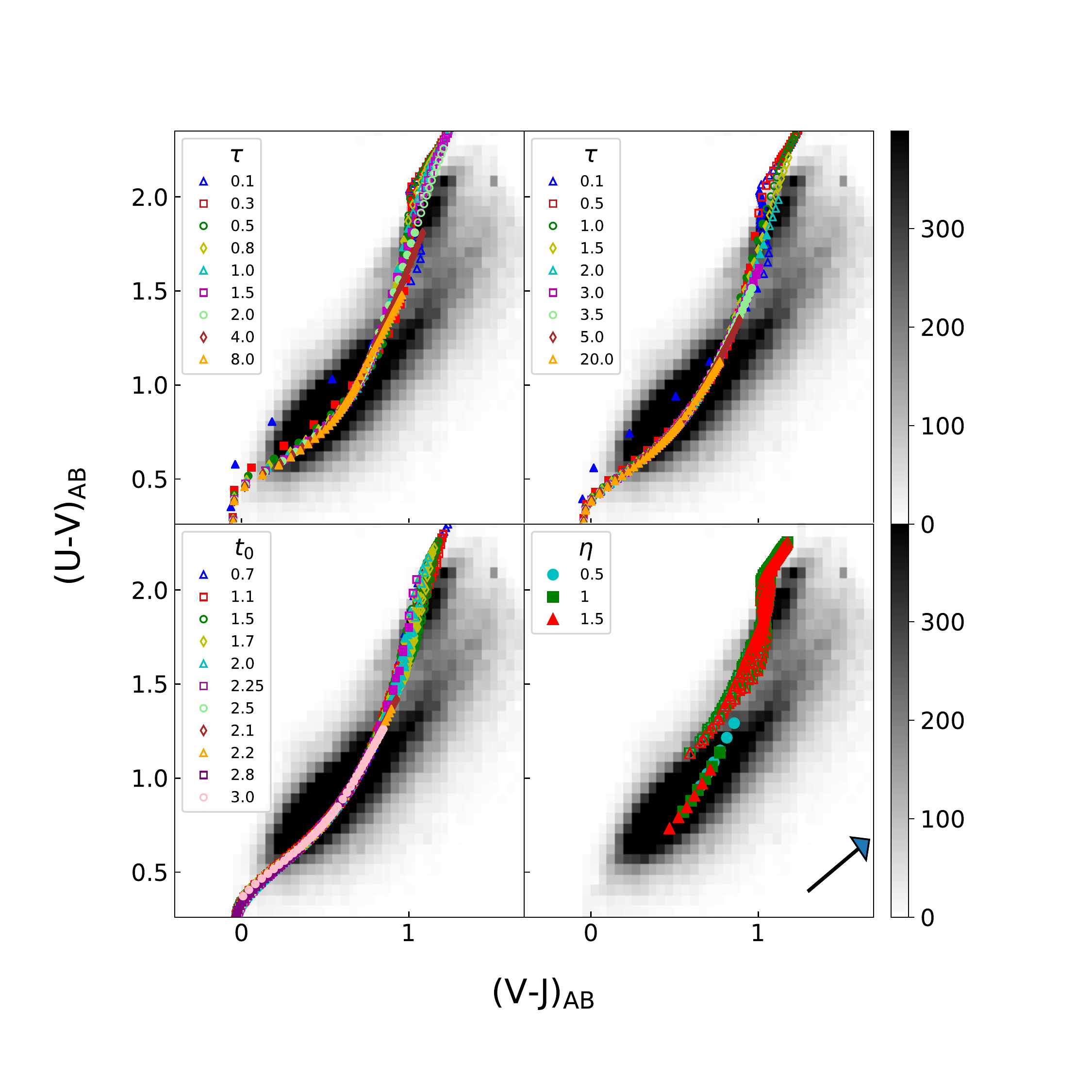}}
\caption{Same as in Fig. \ref{fig_two_cols} but adopting  Maraston SSP models with solar metallicity.}
\label{two_c_cla}
\end{figure*}

\begin{figure*}
\centering
\resizebox{\hsize}{!}{
\includegraphics[angle=0,clip=true]{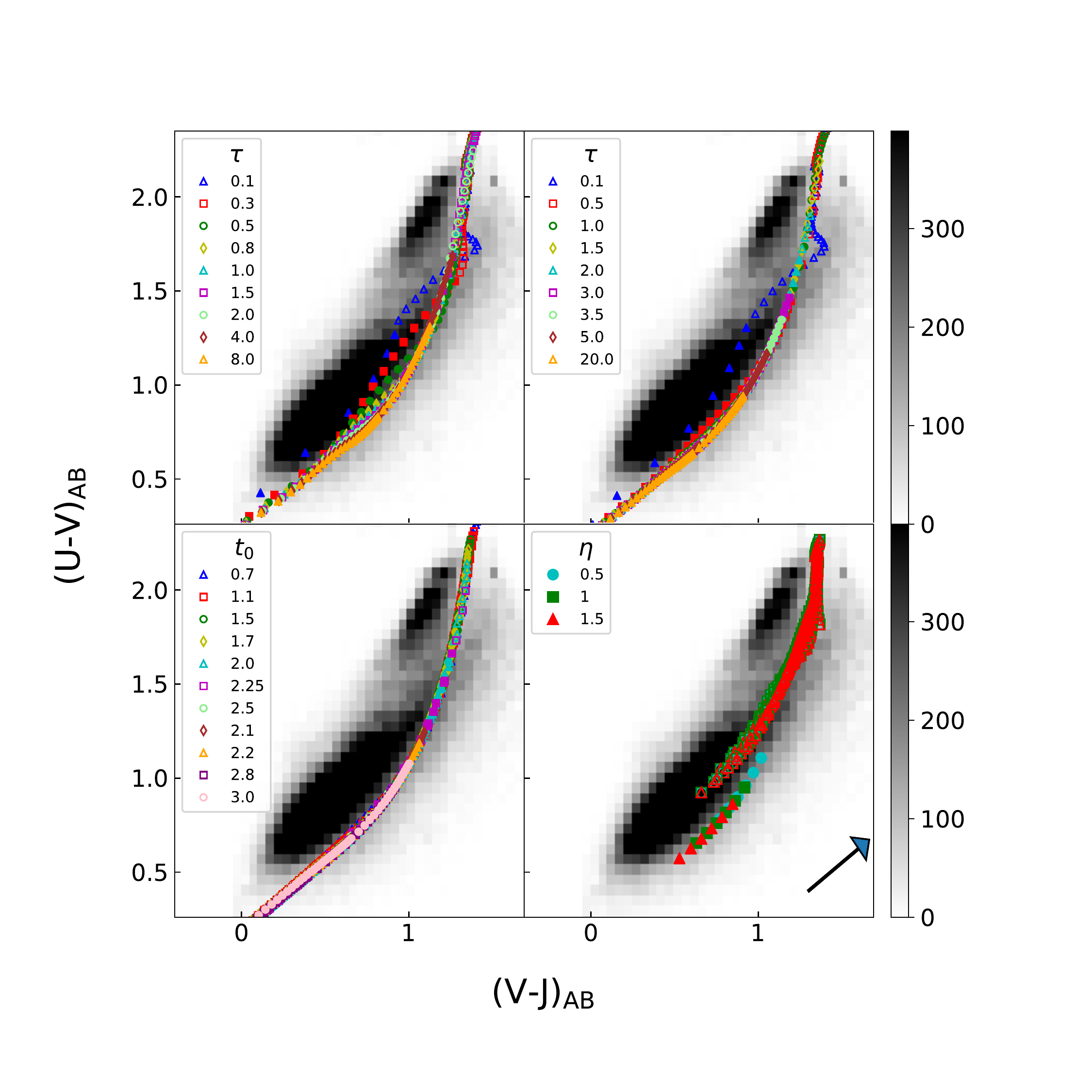}}
\caption{Same as in Fig. \ref{fig_two_cols} but adopting  MG08 SSP models with solar metallicity.}
\label{two_c_leo}
\end{figure*}

Figures \ref{two_c_cla} and \ref{two_c_leo} show, in the same fashion as Fig. \ref{fig_two_cols} in the main text, the models computed with the various SFH prescriptions, respectively adopting the Maraston and the MG08 sets of SSP models. It can be noticed that the shape of  the correlation between the $U-V$ and $V-J$ colours of galaxies with different age distributions depends on the adopted set of  SSPs,  and indeed it reflects the behaviour of pure SSP models on the two-colour diagram. For example, in $V-J$, old solar metallicity SSPs in Maraston are bluer than in BC03, which are bluer than MG08 models; this  trend is reproduced by the colours of passive galaxies in Figs. \ref{two_c_cla}, \ref{fig_two_cols} and \ref{two_c_leo}. The main axis of the distribution of star forming galaxies is not
well matched by Maraston models, which show a too rapid redward evolution of the $U-V$ colour at the red end of the $V-J$ distribution. A similar steep trend is also exhibited by the solar metallicity models based on the MG08 SSPs. 

However, the colours of model galaxies are sensitive to the metallicity, so that the distribution of the data on this diagram could be met with a suitable distribution of chemical composition among the galaxies. Fig.~\ref{fig_cc_z} shows the colours of log-normal SFH models based on the Maraston and on the MG08 sets for subsolar and supersolar metallicities. Using Maraston models, the star forming galaxies could be matched with subsolar metallicity, while the passive galaxies would be better represented with solar abundances. When using MG08 models, the colours of passive galaxies are better matched by the subsolar metallicity set, while the star forming galaxies require a metallicity lower than $\sim 0.4\ Z_\odot$.

Still, all in all, the BC03 solar metallicity models appear to capture in the best way the general features of the galaxies distribution on this plane, especially the slope of the main axis of the distribution.

\begin{figure}
\centering
\resizebox{\hsize}{!}{
\includegraphics[angle=0,clip=true]{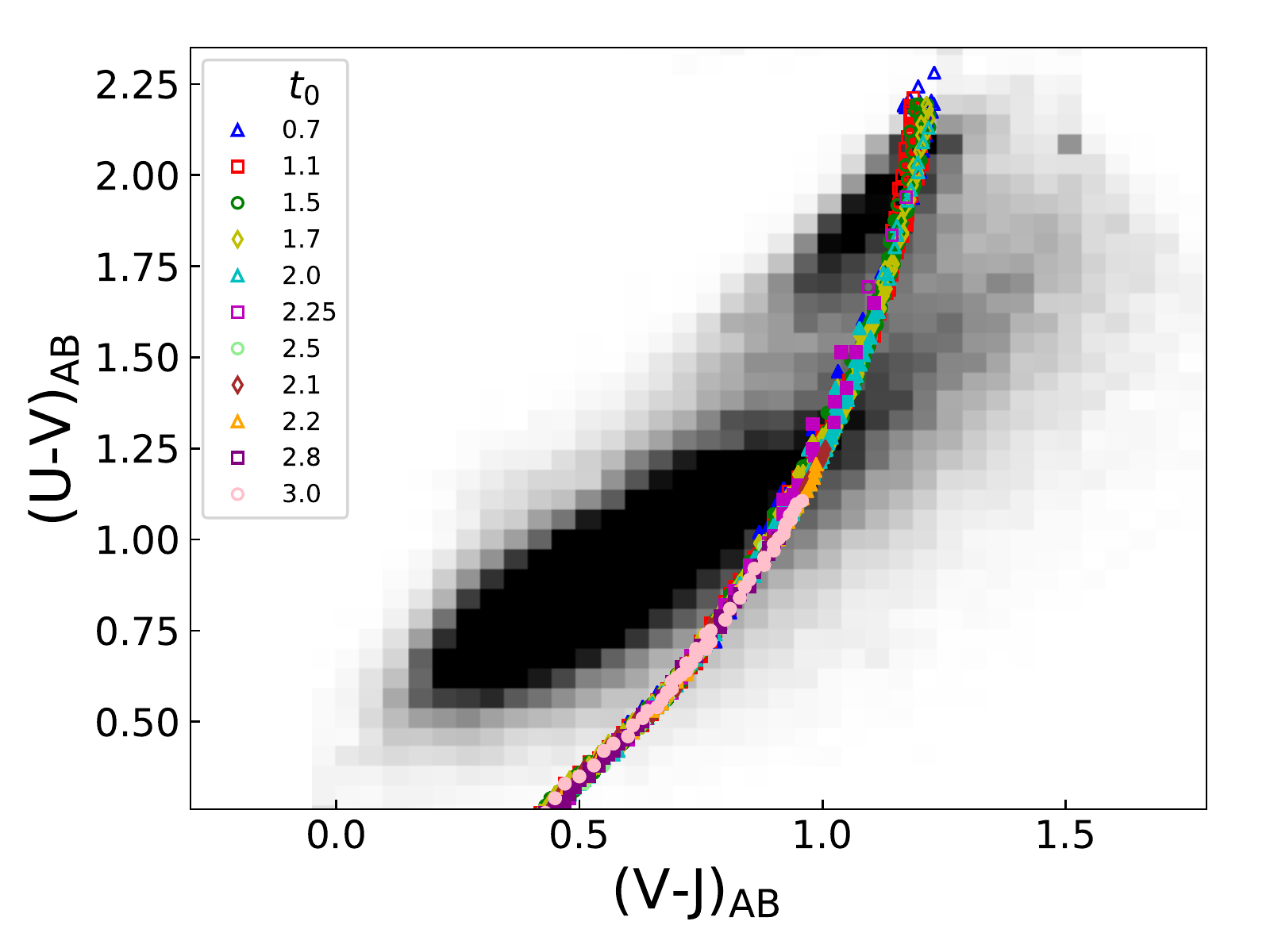}}
\caption{Two-colour diagram for the SUDARE galaxies (greyscale) compared to models computed with the log-normal SFH and the BPASS population synthesis models with solar metallicity}.
\label{two_c_bpass}
\end{figure}

\begin{figure*}
\centering
\resizebox{\hsize}{!}{
\includegraphics[angle=0,clip=true]{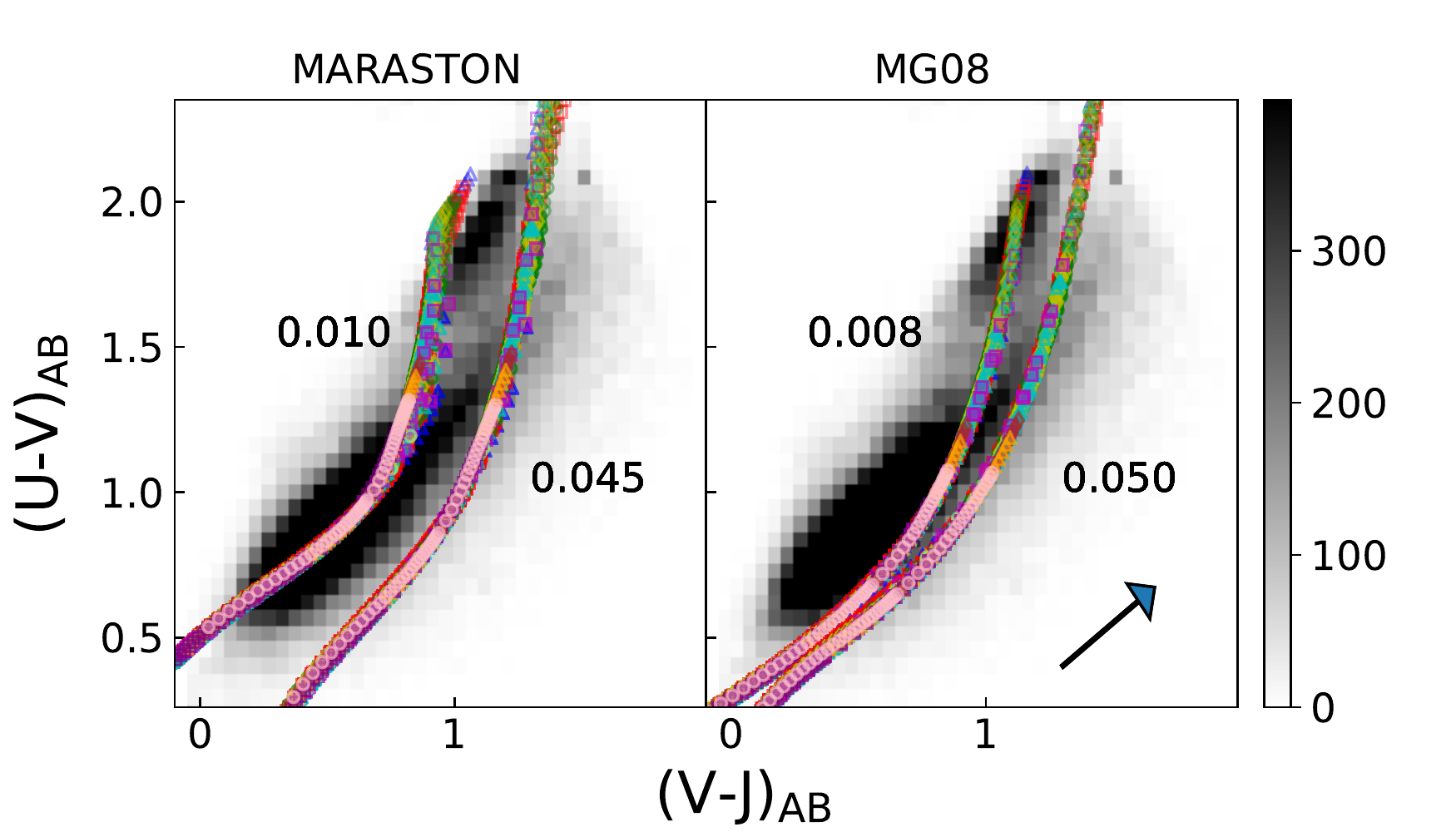}}
 \caption{Two-colour diagram for log-normal SFHs using the Maraston (left panel) and the MG08 (right panel) SSP models with subsolar (upper loci) and supersolar (lower loci) metallicities.
 The models are superimposed on the distribution of the rest-frame colour of the SUDARE galaxy sample. The symbol and colour encoding is the same as in the bottom left panel of Fig. 
 \ref{fig_two_cols}.}
          \label{fig_cc_z}
\end{figure*}

Although we do not aim to contrast the properties of  stellar population models available in the literature, we also computed galaxy models based on stellar populations including binaries. Figure \ref{two_c_bpass} shows model galaxies computed with the log-normal SFH and the colours of solar metallicity, coeval stellar populations including binaries from the BPASS v2.2 suite \citep{Stanway_2018}, with the default IMF. Including binaries does not yield an improved description of the observed galaxy colours. The origin of the discrepancy may be the subject of a future study.
\vfill

\section{Determining the SNIa productivity}  

\begin{figure}
\centering
\resizebox{\hsize}{!}{
\includegraphics[angle=0,clip=true]{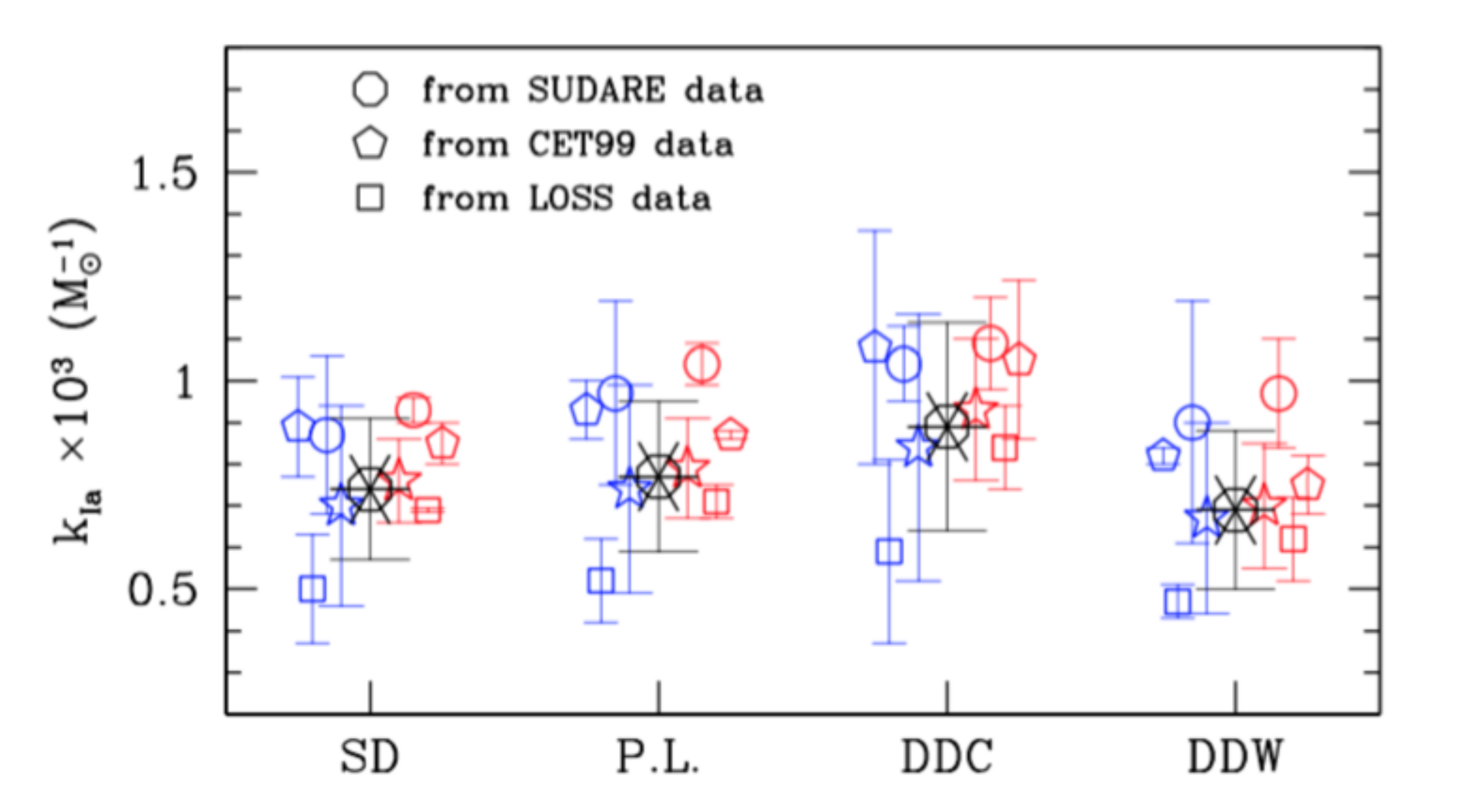}}
\caption{SNIa productivity resulting from the calibration of the models on the rates measured on galaxies with intermediate colour. The point type encodes the observational survey as in the legend. Each of the four groups refers to a different  DTD labelled on the bottom axis. Within each group, points to the left (in blue) result from the calibration on galaxies binned in $U-J$ (as in Fig. \ref{fig_corr_umj}); points to the right  (in red) from the calibration on galaxies binend in $B-K$ (as in Fig. \ref{fig_corr_bmk}). The error bars reflect the statistical uncertainties of the measurements of the rates. For each DTD, the \kia\ values determined from the three surveys are combined to provide a weighted average \kia\ shown as a blue (left) and red (right) star.  Finally, the asterisked circle (black) shows the value of \kia\ obtained combining the determinations from the two colours in a weighted mean. The error bars on the average values show the relative 1$\sigma$ dispersions.}   
\label{fig_kia}
        \end{figure}

By construction, the SNIa rate at late epochs in a system with a flat age distribution is equal to the SNIa productivity (\kia) divided by the age of  the system (see Eq. (\ref{eq_rate_nor})), independent of the DTD.  Stellar populations constructed with a constant SFR for 13 Gyr have $(U-J)_{AB} = 1.8$ and $(B-K)_{AB} =  1$, for BC03  solar metallicity SSP models. Thus the most robust calibration of \kia\ is obtained from the SNIa rate in systems with intermediate colours. In practice, however, galaxies with the same intermediate colour may have  different SNIa rates (see Fig. \ref{fig_corr_umj}) owing to different individual SFHs. Therefore, the rate measured in, for example the colour bin centred at 
$U-J=2$,  to some extent is sensitive to the specific distribution of the SFHs of the galaxies populating the bin. 
This effect can be taken into account only by fitting the age distribution in the individual galaxies, which we deem as a too
refined procedure given the heterogeneous  datasets in hand. We adopt instead a strategy which aims at averaging over a wide parameter space, by considering the model predictions in the two central colour bins for the LOSS and CET99 samples and in the two reddest colour bins for the SUDARE sample. 
Operationally we proceed as follows: we compute all the models shown in Fig. \ref{fig_abramo} with a fixed time step of 0.1 Gyr, mimicking a galaxy population evenly distributed among the parameters values in Fig. \ref{fig_abramo}. Then we compute the average of the theoretical rate (for \kia=1) of all models which fall in a specific colour bin. The ratio of the rate measured in the colour bin  to the computed average model rate yields a value for \kia. We repeat the procedure in the adjacent colour bin to construct an average value for \kia\ which best represents the measured rate in the galaxies with intermediate colour. We perform this computation for each DTD and each observational sample and show the results  in Fig.  \ref{fig_kia}. The figure shows that the various determinations are in broad agreement.  Compared to LOSS, the SUDARE and CET99  datasets yield systematically higher values of the productivity in both colours, but the discrepancy is lower when calibrating on the rate versus $B-K$ correlation. For each DTD, the average values of  \kia\ derived from the two colours are in excellent agreement and the dependence of the final average values for the different DTDs is negligible. These features support the conclusion that the derived value of the productivity is very robust. 

\end{appendix}

\end{document}